# Bifurcation and chaos for a new model of trigonal centrifugal governor with nonsmooth control

**Yanwei Han** (✉)
School of Civil Engineering, Henan University of Science and Technology, Luoyang 471023, China
e-mail: yanweihan@hotmail.com
**Zijian Zhang**
School of Astronautics, Nanjing University of Aeronautics and Astronautics, Nanjing 210016, China

**[Abstract]** The flywheel ball and hexagonal structures in the design of the classical centrifugal governor systems lead to both modeling and analytical difficulties. In the present paper, a new trigonal centrifugal governor (TCG) is proposed in an attempt to overcome both of these difficulties by introducing the radical nonlinearity and nonsmooth control strategy in the simple and clear formula. The nonlinear dynamical behaviors of this new model are investigated for both the autonomous and the non-autonomous cases. The three equations of motion of the TCG are presented based on Euler-Lagrange's equation and the theorem of angular momentum. The velocity, nonlinear restoring force and nonsmooth torque surfaces are plotted to display the complex relationship of parameter change dependence. Secondely, the equilibrium bifurcation and stability analysis for autonomous system are investigated to show the pitchork bifurcation phenomena and the saddle-focus point respectively. It is found that the system bears significant similarities to the Duffing system with mono-stable and bi-stable characteristic. Finally, the three-dimensional Melnikov method is defined and employed to obtained the analytical chaotic thresholds for non-autonomous centrifugal governor system, and numerical results of chaotic behaviors verify the proposed theoretical criteria of the non-autonomous system. The experimental studies are carried out to validate the theoretical and numerical results.

## 1. Introduction

The mechanical-centrifugal speed governor plays an important role in many rotating machinery such as hydraulic turbine, diesel engine, steam engine, gasoline engine, automobile engine and so on. It is a device that automatically avoids damage or failure caused by sudden changes of load torque by using controls rotational speed of rotational machines. In the year 1788, James Watt has constructed the first Watt steam engine and initiated the revolutionary power-generating machine. Thomas Mead invented a patent of the centrifugal of fly-ball governor to regulate the action of windmills. Maxwell and Vyshnegradskiistudied on the local stability of the Watt governor system were reported in 1868 and 1876 respectively [1,2]. Pontryagin presented a local stability of a simplified version of the Watt centrifugal governor system base on Vyshnegradskii work [3]. This venerable governor device remains in public eyes as icon of feedback control and automatic devices. Nowadays it used extensively as paradigm in dynamic control of stability, chaos, synchronization and Hopf bifurcation [4-6].

The general basis of stability theory established by Lyapunovin 1892 widely applied in mathematics, mechanics and control fields. Presently many scholars use this theory to analyze the stability of centrifugal governor. Denny analyzed the stability of the fly-ball governor anew and generalized to arbitrary governor geometry [7]. Sotomayor et al derived sufficient conditions of the equilibrium stability for the hexagonal centrifugal governor system based on the analysis of the Lyapunov stability and the Hopf bifurcation [8]. Luo et al investigated the dynamic behaviors of optimality and the accelerated adaptive stabilization with controller of speed function for the fractional order governor system [9]. Deng et al considered the stability and bifurcation of the equilibrium for a hexagonal centrifugal governor system with and without the effect of time-delay [10]. Alidousti et al investigated the stability and bifurcation of equilibrium and parametric condition of Hopf bifurcation for a fractional order governor system[11]. Kuznetsov et al presented a mathematical model of Sayano-Shushenskaya hydropower plant to explain the stability and accident vibrations [12]. On the other hand, Hopf bifurcation is a local stable phenomena describing the creation of limit cycles near a fixed point, example include problem from physics, chemistry, biology, electron, and others which has attracted interesting of more researchers. Zhang et al studied Hofp bifurcationin and hyperchaotic motion of an hexagonal centrifugal governor system with spring [13]. Wen et al proposed a feedback control strategy to design the amplitude and frequency of Hopf bifurcation for an hexagonal centrifugal governor system [14].

The chaotic behavior by a set of three ordinary differential equations discovered by Lorenz in 1963 accidentally [15]. The serious problem of the centrifugal governor system caused by the chaotic oscillation and unpredictable and irregular behaviors has been catching many researcher's interesting. Zhu et al derived the condition of Melnikov function for chaotic vibration of a one degree of freedom simplified governor model undergoes a harmonic variation [16]. Ge et al proposed two produces to control chaos synchronization of a rotational machine with hexagonal centrifugal governor [17]. Chu et al presented chaotic attractor, Hopf bifurcation and bifurcation diagram of a centrifugal governor system and applied the feedback control method to control the chaos synchronization [18]. Aghababa et al proposed a adaptive controller for stabilization of chaotic motion of centrifugal governor systems with in limited time [19]. Rao et al studied nonlinear dynamics of periodicity and chaos and fractal characteristic of the self-similarity structure and Arnold tongues for mechanical centrifugal governor [20]. Luo et al designed a chaos controller of tangent barrier Lyapunov function to understand chaotic nature of centrifugal flywheel governor system [21]. Yan et al investigated the multi-stable dynamical behaviors of integer- and fractional-order governor system when stochastic noise is considered by coexisting chaotic attractors and Lyapunov exponents [22]. Meanwhile, in 1963 Melnikov investigated the trajectories of the perturbed time-periodic Hamiltonian systems of two-dimensional vector fields and providing analytical method for determine whether transverse homoclinic orbits to hyperbolic chaotic motions [23]. The Melnikov's function have been demonstrated to be a useful tool for the chaotic dynamics. Lin employed a function space approach to obtain bifurcation function near heteroclinic or homoclinic cycles and shown a examples for generalisations of Silnikov's theorems [24]. Du et al established higher-order Melnikov's method to discussed homoclinic bifurcation of piecewise linear system of an inverted pendulum with impact under external periodic excitation [25]. Chacon presented review summarizes of Melnikov's method in non-autonomous oscillator systems with homoclinic and heteroclinic chaos [26]. Awrejcewicz et al studied chaotic behavior of a rotated Froude pendulum with Coulomb-type friction, by analytical approach of Melinkov 's technique to a homoclinic bifurcation [27]. Yin et al investigated the chaotic threshold of Melnikov method for the nonlinear Schrodinger equation under periodic perturbation [28].

Therefore, many authors have steadily and systematically improved the the knowledge of two type of centrifugal governor of the flywheel ball and the hexagonal. In our study, this paper present novel trigonal centrifugal governor system with oblique springs to overcome the modeling and analytical difficulties, this model simple enough to allow for straightforward analytical results. This paper is organized in the following way. In section 2 the modeling and dynamical equations of centrifugal governor are obtained by Euler-Lagrange's method and moment of momentum theorem. In section 3 the equilibria of bifurcation and stability of saddle focus type are analyzed to show chaotic feature of the autonomous system. In section 4, the three dimensional Melnikov's function is defined to reveal parameter condition of chaotic boundaries for the nonautonomous system. The experiment method is carried out for model validation. The concluding remarks are given In section 6.

## 2. Modeling of trigonal centrifugal governor

### 2.1 Equation of motion for governor system

The basic Watt governor is illustrated in Fig. 1(a), the feedback control of engine system relies on the rotation speed of fly-balls which rotating around shaft. The rotating velocity of governor system leads to the height of the sleeve depending on the centrifugal force. The displacement of sleeve controls the flow of working fluid by throttle valve by mean of the rod, maintaining steady rotational speed of engine system [5]. In Fig. 1(b), consider a nonlinear mechanical model of trigonal centrifugal governor system with a pair of oblique springs. This system comprised of two lumped masses $m$, moving horizontally on the slide bar and linked by a pair of oblique linear elastic springs of stuffiness $k$ and original length $b$ of the springs in the equilibrium position which are pined to rigid supports. These two springs are assume to remain straight under stretchable or compressive loading. Two basic assumptions for this governor system as follows: (i) The mass of the wirerope, rod and springs so small that can be ignored; (ii) The viscous damping between the slide rail and two masses is constant $c$.

As notations of the centrifugal governor modelling are shown schematically in Fig. 1(b), the kinetic energy ($KE$), potential energy ($PE$) and Rayleigh function $\Psi$ can be written as

$$KE = m\dot{x}^2 + m\omega^2 x^2 + 0.5I\dot{\theta}^2 \tag{1}$$

$$PE = k\left((x^2+a^2)^{0.5} - b\right)^2 \tag{2}$$

$$\Psi = c_1\dot{x}^2 + 0.5c_2\dot{\theta}^2 \tag{3}$$

where $m$ (kg) is the mass of the sliding block, $x$ (m) is the distance between the mass and the rotational axis , $\dot{x} = \mathrm{d}x/\mathrm{d}t$ (m/s) is the velocity of mass, $\dot{\theta} = \omega$ (rad/s) is the rotational speed, $a$ (m) is vertical distance of the simply supported for the coil springs, $b$ (m) is represent the free length of the coil springs, $c_1$ (m·s/m) is the damping coefficient between fly mass and bar, $c_2$ (N·m·s/rad) is the torsional damping coefficient between shaft and bearing, and $t$ (s) is the time .

(a) (b)

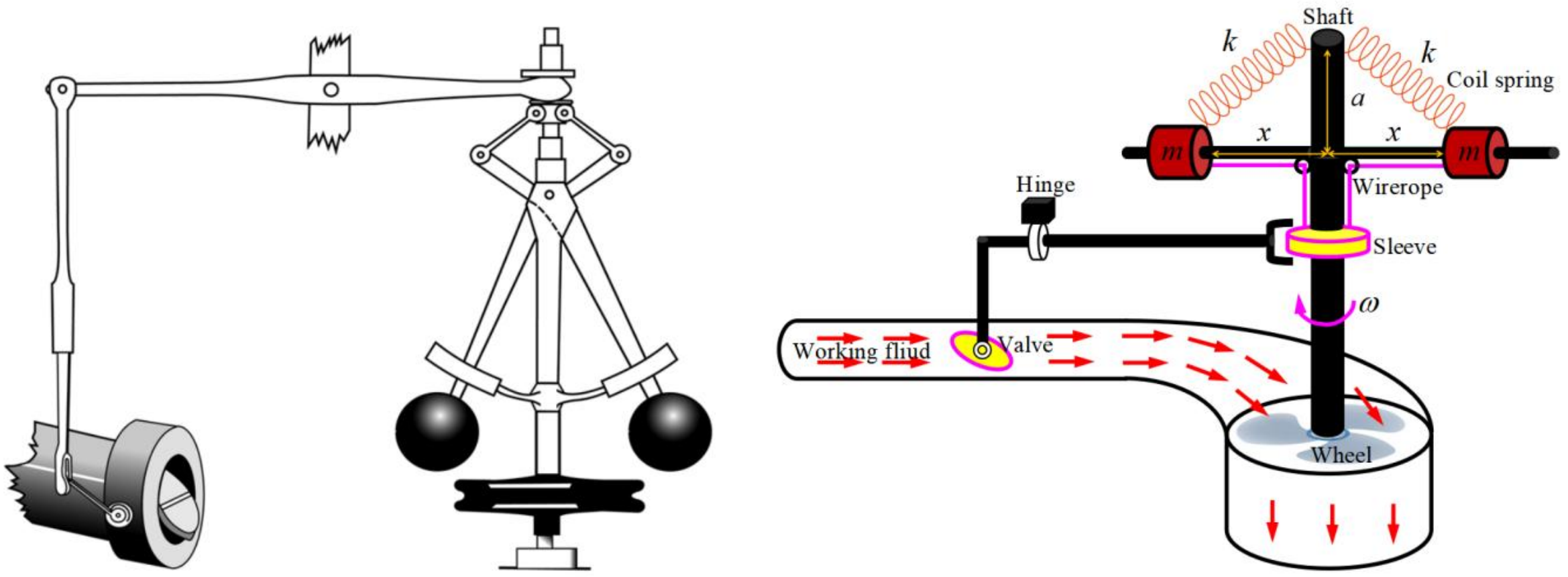

Fig. 1 The centrifugal governor. (a) Physical model with flywheel and throttle valve of Watt's [5] steam engine. (b) New governor model attached by a pair of masses in triangle construction [29].

The Lagrange's function is written as follows

$$\Pi = KE - PE. \tag{4}$$

Euler-Lagrange's equations can be written as

$$\begin{cases} \dfrac{\mathrm{d}}{\mathrm{dt}}\left(\dfrac{\partial \Pi}{\partial \dot{x}}\right) - \dfrac{\partial \Pi}{\partial x} + \dfrac{\partial \Psi}{\partial \dot{x}} = Q_1 \\ \dfrac{\mathrm{d}}{\mathrm{dt}}\left(\dfrac{\partial \Pi}{\partial \dot{\theta}}\right) - \dfrac{\partial \Pi}{\partial \theta} + \dfrac{\partial \Psi}{\partial \dot{\theta}} = Q_2 \end{cases} \tag{5}$$

Straight forward calculation from Eq. (5), the equation of motion of this TCG system with trigonal configuration is therefore given below

$$\begin{cases} m\ddot{x} + c_1\dot{x} + kx(1 - b/(x^2+a^2)^{0.5}) - m\omega^2 x = Q_1, \\ I\dot{\omega} + c_2\omega = Q_2, \end{cases} \tag{6}$$

where the generalized forces $Q_1$=0 and $Q_2$=$M_1$−$M_2$, the Eq. (6) is the governing equations of motion for the governor.

For this rotating system, the input torque $M_1$ (N·m) produced by the engine steam, fuel oil or water. The load torque $M_2$ (N·m) caused by a change in load. Applying the moment of momentum theorem, the equation of rotational motion may be expressed as

$$I\dot{\omega}+c_2\omega-M_1+M_2=0, \tag{7}$$

where $I$ (kg·m$^2$) is the moment of inertia for this centrifugal governor.

For a given rotating speed $\omega_0$, the average engine torque $\bar{M}_1$ (N·m) is introduced to substitute of the input torque $M_1$. The height of the valve controls the work fuel depends on the displacement $x$. Thus, the input torque $M_1$ can be expressed as

$$M_1=\bar{M}_1+h(|x_0|-|x|), \tag{8}$$

where $x_0$ (m) represents the displacement corresponding to $\omega_0$ (rad/s) and $h$ (N) demote a constant of proportionality standing for produced by the decrease for steam with the displacement $x$.

For the rotational motion of the TCG system, make using the theorem of angular momentum the angular velocity equation of motion is given by

$$I\dot{\omega}+c_2\omega+h|x|=M \tag{9}$$

where $M=h\,|\,x_0\,|+\bar{M}_1-M_2$ is the equivalent load torque. The Eq. (9) is the second governing equation of motion for TCG model.

From Eqs. (6) and (9), the three-dimensional form differential equations of motion of TCG system can be written in following

$$\begin{cases}\dot{x}=y,\\ \dot{y}=-(c\dot{x}+kX(1-b/(x^2+a^2)^{0.5}))/m+\omega^2 x,\\ \dot{\omega}=(c_2\omega-h\,|\,x\,|+M)/I.\end{cases} \tag{10}$$

In order to conveniently study the dependence of the system on parameters, we make use of the following non-dimensional transformation formulas in the variables, parameters and time. The scaling formulas for the governing equations (10) can be expresses as

$$\begin{gathered}X=x/b, Z=\omega/\omega_n, \omega_n^2=k/m, T=\omega_n t, \alpha=a/b,\\ \xi_1=c_1/(km)^{0.5}, \xi_2=c/(\omega_n I), \gamma=hb/(I\omega_n^2), \delta=M/(I\omega_n^2),\end{gathered} \tag{11}$$

where $X$ and $Z$ are the dimensionless displacement and angular speed respectively, $\omega_n$ is the natural frequency, $T$ is the dimensionless time, $\xi>0$ is the damping coefficient, $\alpha$ is the dimensionless geometry parameters, $\gamma>0$ is the dimensionless proportional constant, $\delta>0$ is the dimensionless torque of the load.

Using the above scaling transformation formulas of Eq. (11), Eq. (10) can be reducing to the three nondimensional differential equations of the autonomous centrifugal governor system. Introducing the notation $Y$ for the velocity $X'$, one can get the three-order differential equation of this governor as follows:

$$\begin{cases}X'=Y,\\ Y'=-\xi_1 X'-X(1-1/(X^2+\alpha^2)^{0.5})+Z^2X,\\ Z'=-\xi_2 Z-\gamma\,|\,X\,|+\delta.\end{cases} \tag{12a}$$

where the primes denote with respect to the dimensionless time $T$.

Leeting $V_X=-Y$ the nondimensional linear velocity, $F_Y=-(-\xi_1 X'-X(1-1/(X^2+\alpha^2)^{0.5})+Z^2X)$ the dimensionless restoring force with fractional and rational nonlinearities and $M_Z=-(\xi_2 Z-\gamma\,|\,X\,|+\delta)$ the dimensionless torque, Eq. (12a) can be rewritten as

$$\begin{cases}X'=-V_X,\\ Y'=-F_Y,\\ Z'=-M_Z.\end{cases} \tag{12b}$$

The main difficulty arises from fact that this Eq. (12) involves both fractional and radical non-inearities due to the geometrical configuration features of oblique springs. Those new-type nonlinearities can not solved by the analytic theory of nonlinear vibration base on the Taylor's series theorem for its approximation and truncation[30].

## 2.2 The nonlinear force and nonsmooth control torque

For $Z=$ constant and $\xi=0$, the autonomous governor system (12) can be translated to the following system in the form of two first-order differential equations

$$\begin{cases} X'=Y, \\ Y'=-X(1-1/(X^2+\alpha^2)^{0.5}))+Z^2X, \end{cases} \tag{13}$$

the Eq. (13) defined as unperturbed centrifugal governor system.

The nondimensionlal velocity and nonlinear rational restoring force for the governor system (13) are defined respectively as

$$\begin{cases} V_X=-Y, \\ F_Y=X(1-1/(X^2+\alpha^2)^{0.5})-Z^2X. \end{cases} \tag{14}$$

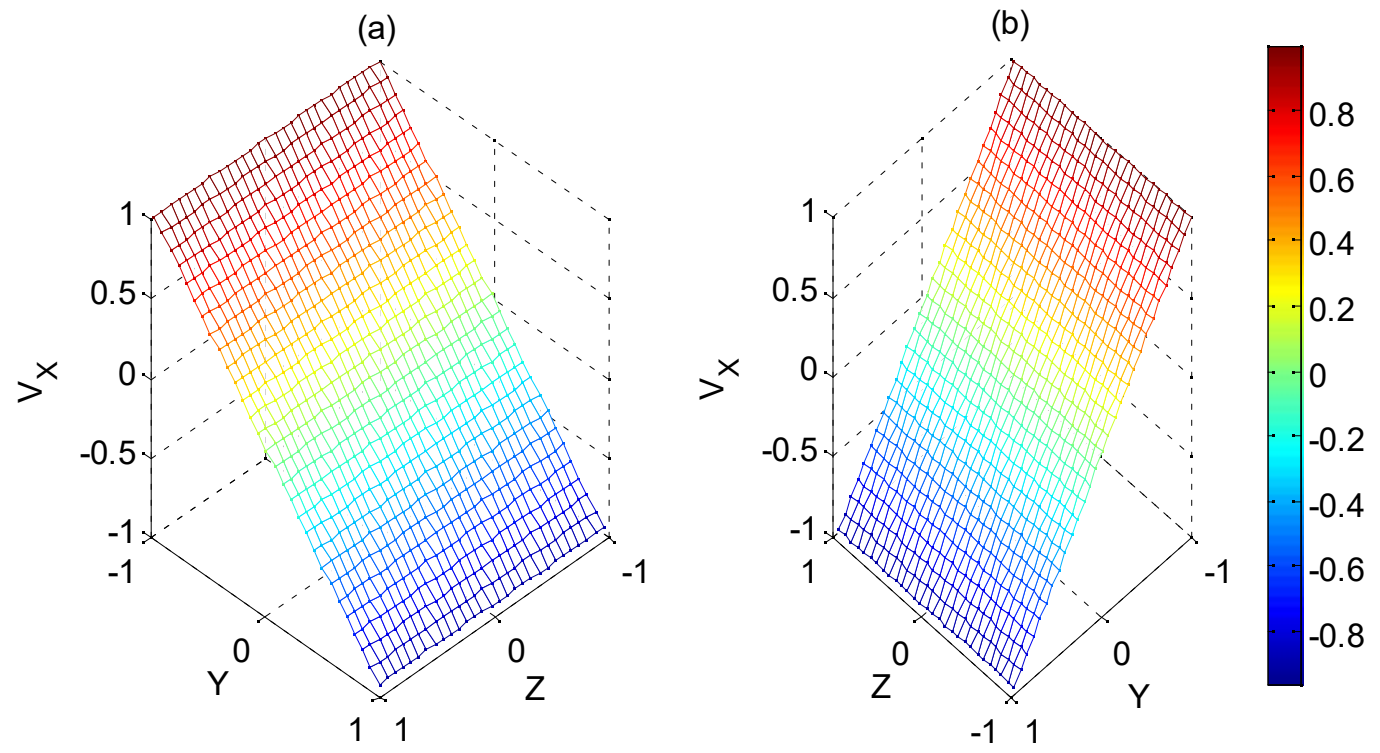

Fig. 2 The non-dimensional velocity function $F_X$ surfaces in space ( $Y$, $Z$, $V_X$) for the Eq.(14). (a) In the (1, 1, 1) view. (b) In the (−1, 1, 1) view.

As shown in Fig. 2, the velocity surfaces of $V_X$ versus $X$ and $Z$ are plotted. It is easy to see that the force function $V_X$ is a flat plane, which mean that the speed variable $V_X$ is negatively proportional to $Y$ and independent of $Z$.

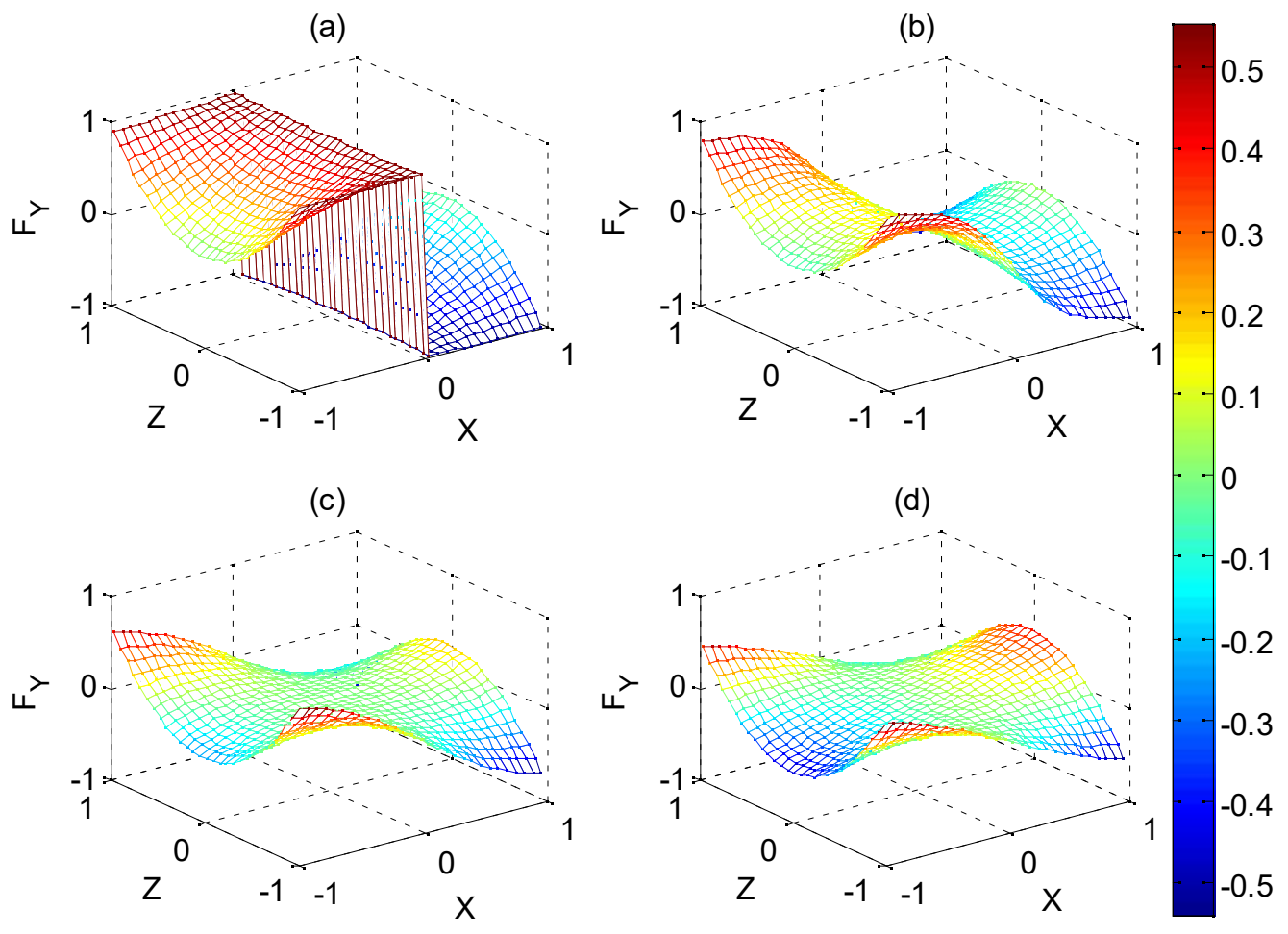

Fig. 3 The non-dimensional force $F_Y$ surfaces in space ( $X$, $Z$, $F_Y$) for the restoring Eq. (14). (a) $\alpha=0.0$, (b) $\alpha=0.5$, (c) $\alpha=1.0$, (d) $\alpha=1.5$.

The complex resistance force $F_Y$ surface as the function of $X$ and $Z$ for the different value of parameter $\alpha=0$, 0.5, 1.0 and 1.5 shown in Fig. 3 respectively, form which we see the clearly the change trend of the nonlinear force for the variation of position parameters $X$ and angular velocity parameter $Z$. It is most interesting that the discontinuous characteristic in section of $X$=0 in Fig. 3(a), at which the infinity derivative of stiffness $K(X=0)=\partial F_X/\partial X=\infty$ with nonsmooth bistable characteristic. Fig. 3(b) gives the smooth double-well dynamics for $\alpha$ = 0.5. Fig. 3(c) shows the only single-well nonlinear force curves for $\alpha$=1.0. Fig.3 (d) shows the only single-well nonlinear force curves for $\alpha$ =1.5.

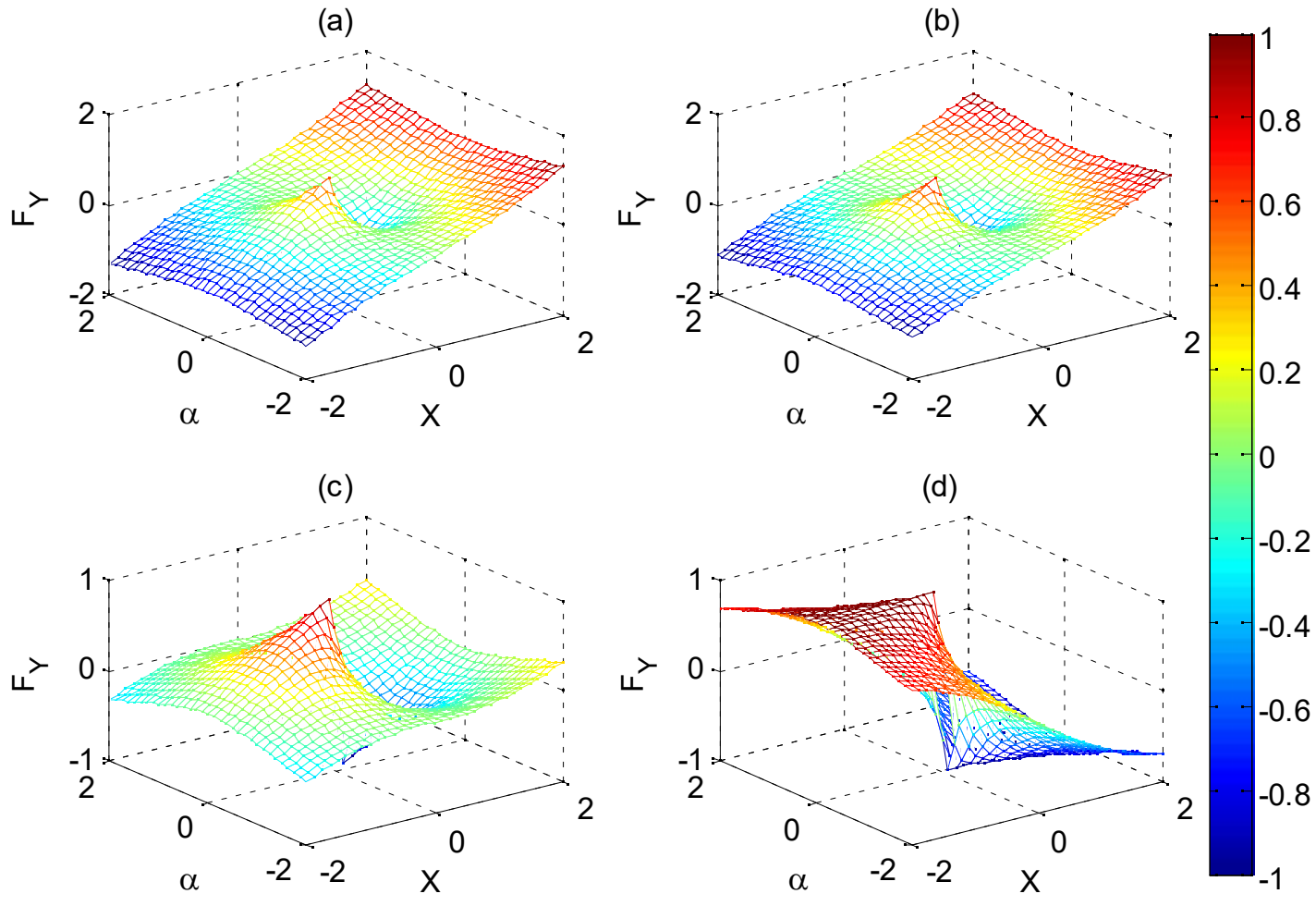

Fig. 4 The dimensionless restoring force $F_Y$ characteristic surfaces in space ($X$, $\alpha$, $F_Y$) for the restoring force Eq. (12). (a) $Z$=0.0, (b) $Z$=0.3, (c) $Z$=0.7, (c) $Z$ =1.0.

In Fig. 4, Nonlinear restoring forces of Eq. (14) in space ($X$, $\alpha$, $F_Y$) are plotted for the different value of rotation parameter $Z$ = 0, 0.3, 0.7, 1.0, as a function of both nondimensional displacement $X$ and geometrical ratio $\alpha$. The gradient of the dimensionless restoring force $F_Y$ against $X$ is the stiffness of the TCG system (13) for different value of geometrical ratio $\alpha$. It should be noted that infinite negative stiffness $K = \partial F_X / \partial X = \infty$ characteristic at point ($X$, $\alpha$, $F_Y$)=(0, 0, 0). It is also found that the system (13) exhibited the smooth single-well ($\alpha$>1), smooth double-well (0<$\alpha$<1) dynamics nonlinear force curves and discontinuous double ($\alpha$=1) behavior.

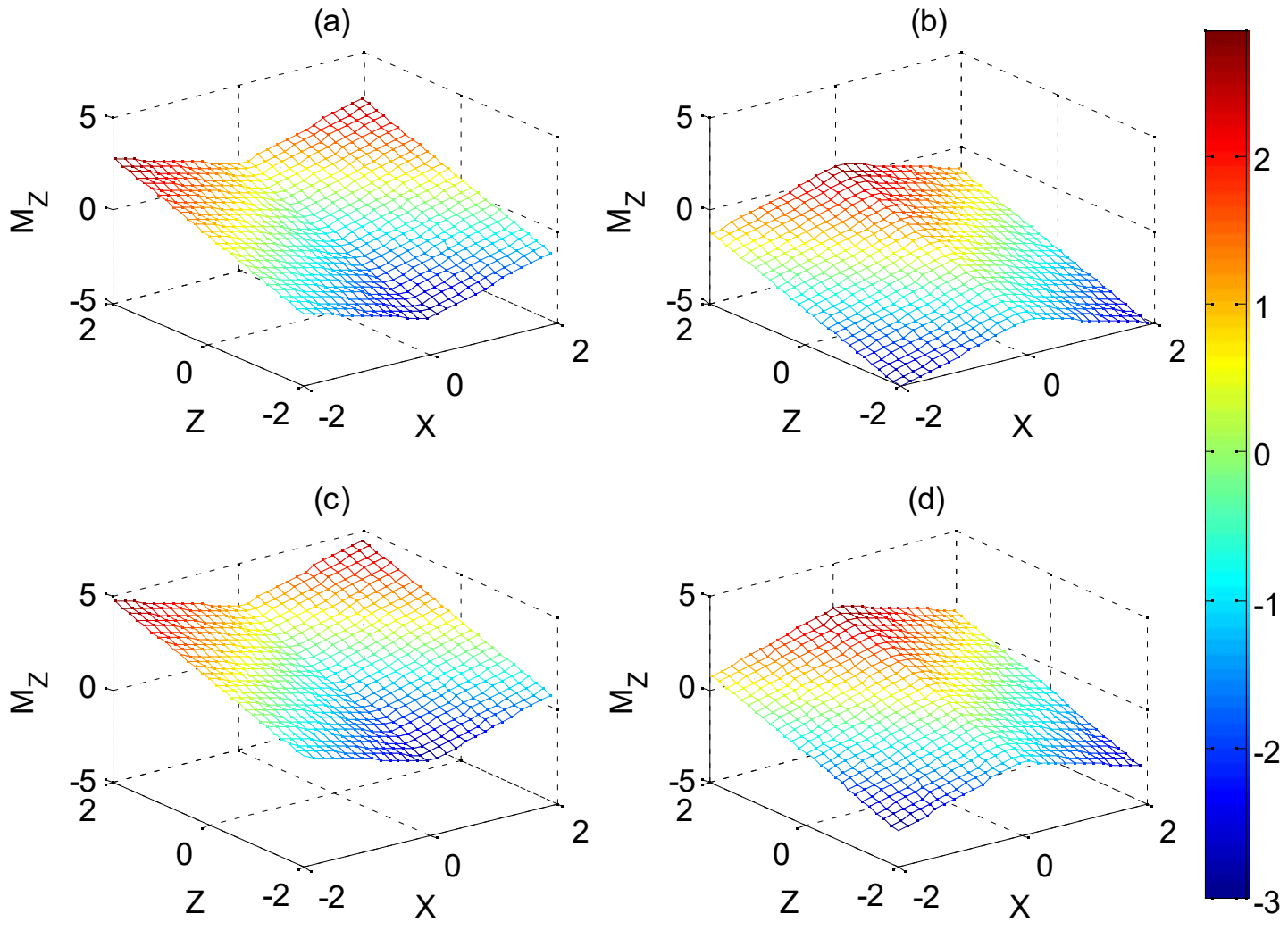

Fig. 5 The dimensionless rotation control torque $M_Z$ surfaces with nonsmooth characteristic in ($Z$, $X$, $M_Z$) space for the Eq. (12). (a) $\gamma$ =1, $\delta$ =1, $\xi_2$=1, (b) $\gamma = -1$, $\delta$ =1, $\xi_2$=1,(c) $\gamma$ =1, $\delta = -1$, $\xi_2$=1, (c) $\gamma = -1$, $\delta = -1$, $\xi_2$=1.

For control torque of rotational function $M_Z$ is written as follow form

$$M_Z = -(\xi_2 Z - \gamma \,|\, X \,|\, + \delta). \tag{15}$$

The rotational control torque function $M_Z$ versus $Z$ and $X$ are illustrated in Fig. 5. In Figs. 5(a) and (c) the control torque of force with V-shaped are plotted in parameter ($Z$, $X$, $M_Z$) space. Similarly in Figs. 5(b) and (d), the torque surface with Λ-shaped given. In Figs. 5(b) and (c), the torque $M_z$ is always greater than or less than zero, which means that the angular velocity keeps increasing or decreasing, therefore these two control strategies is not feasible. As shown in Fig. 5(d), the in the vicinity of zero speed cannot rise revolution, that is, to say this system cannot start, this parameters setting is not feasible. In summary, only the control torque as shown in Fig. 5(a) is reasonable.

## 3. The autonomous system

### 3.1 The equilibrium bifurcation

In this section, we discuss the autonomous dynamics by investigating the equilibrium stability and bifurcations. The equilibrium set of the three equilibria of the unperturbed system (13) without damping and angular speed can be easily obtained and these are

$$\boldsymbol{E}_{0i}=\{X_i|F_Y(X_i,\alpha,Z)|=0\}, \tag{16}$$

where $i=1,2,3$ , $X_1=-\sqrt{1/(1-Z^2)^2-\alpha^2}, X_2=0, X_3=\sqrt{1/(1-Z^2)^2-\alpha^2}$ are three equilibria displacement. The $\boldsymbol{E}_{01}$ and $\boldsymbol{E}_{02}$ are the stable equilibrium position for $\partial F_Y/\partial X(X=E_{01},E_{03})>0$, and the $\boldsymbol{E}_{02}$ are the unstable equilibrium position for $\partial F_Y/\partial X(X=E_{02})<0$ according to the Langrange-Dirichlet theorem.

Fig. 6(a) gives the equilibrium bifurcation surface $\boldsymbol{E}_{0i}$ of Eq. (16) in parameter ($X$, $\alpha$, $Z$) control space for the dimensionless restoring force $F_Y$=0. For $\alpha > 0$, at two equilibrium points $\boldsymbol{E}_{01}$ and $\boldsymbol{E}_{03}$, eigenvalues of the system (13) can be obtained by examining the Jacobian maxtrix of the unperturbed system and they are $\lambda_{11,12}=\lambda_{31,32}=\pm i(1/(1-Z^2)^2-\alpha^2)^{0.5}$, which means that those equilibrium is centers points. In the same way, the eigenvalues of the equilibrium point $\boldsymbol{E}_{02}$ can be obtained as $\lambda_{21,22}=\pm((1-\alpha)/\alpha+Z^2)^{0.5}$ the center points.

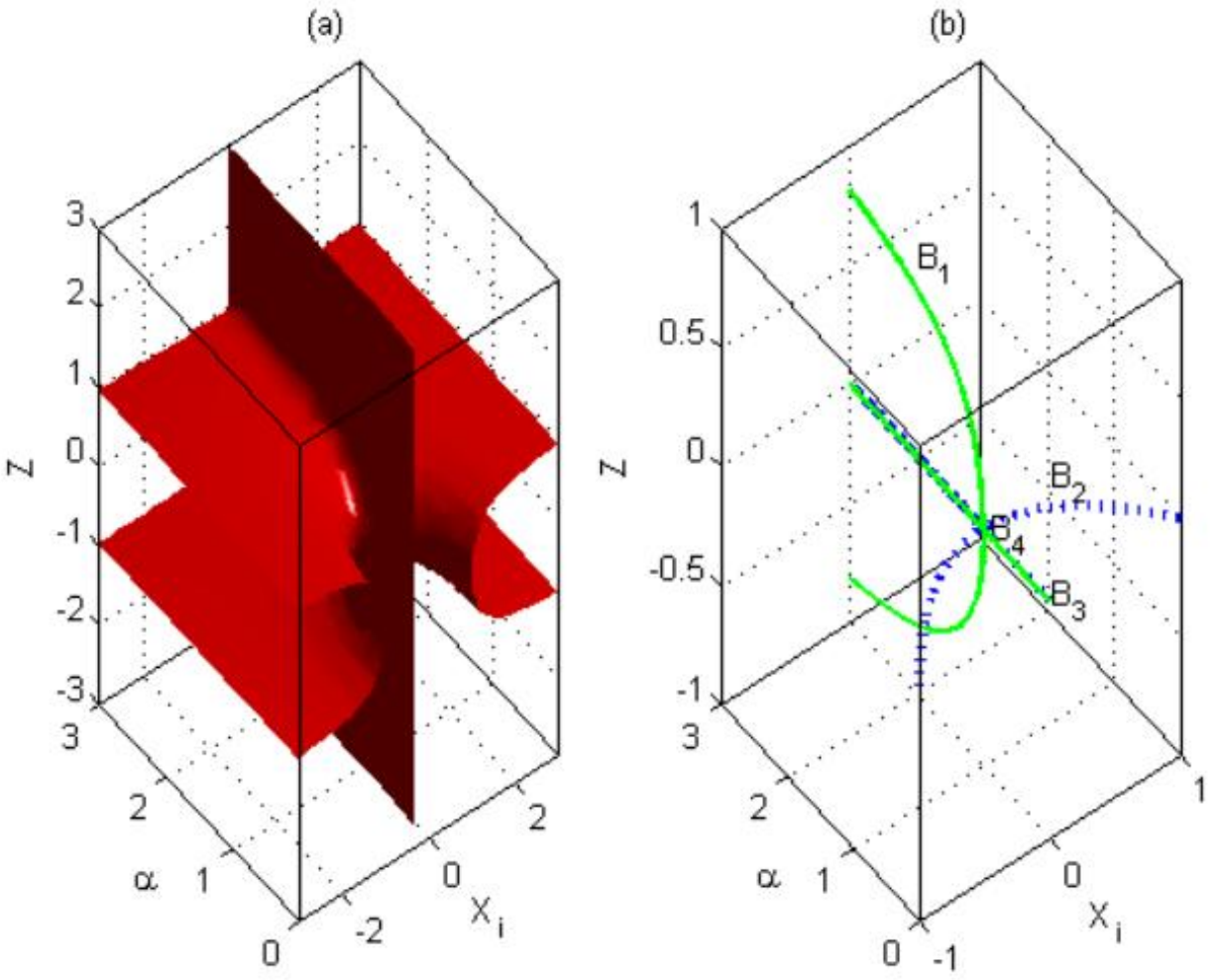

Fig. 6. Equilibrium points surface $\boldsymbol{E}_{0i}$ by the Eq. (16) and bifurcation curves $\boldsymbol{B}_1$ and $\boldsymbol{B}_2$ of Eq. (17). and (a) In parameters space ($X$, $\alpha$, $Z$). (b) Bifurcation sets $\boldsymbol{B}_1$ (green solid curve), $\boldsymbol{B}_2$ (blue dotted curve) and $\boldsymbol{B}_3$ .

In order to examine the influence of non-dimensional geometry parameter $\alpha$ and angular velocity $Z$ on the stability of equilibrium, the bifurcation sets are defined as following

$$\begin{cases}\boldsymbol{B}_1=\{(X,\alpha,Z)|F=0,\partial F/\partial X=0\},\\ \boldsymbol{B}_2=\{(X,\alpha,Z)|F=0,\partial F/\partial Z=0\}.\end{cases} \tag{17}$$

The equilibrium bifurcation sets can derived as in following form

$$\begin{cases}\boldsymbol{B}_1=\{(0,\alpha,Z)|Z(1-1/\alpha+Z^2)=0\},\\ \boldsymbol{B}_2=\{(X,\alpha,0)|X\left(1-1/(X^2+\alpha^2)^{0.5}\right)=0\}.\end{cases} \tag{18}$$

The $\boldsymbol{B}_1$ and $\boldsymbol{B}_2$ are plotted in Fig. 3 (b) denoted by solid and dotted curves, respectively.

The intersection of set $\boldsymbol{B}_1$ and set $\boldsymbol{B}_2$ is given by

$$\boldsymbol{B}_3=\boldsymbol{B}_1\cap\boldsymbol{B}_2=(0,\alpha,0), \tag{19}$$

where $\boldsymbol{B}_3$ is line as shown in Fig. 6(b).

The singular equilibrium point set $\boldsymbol{B}_4$ is defined as

$$\boldsymbol{B}_4 = \{(X,\alpha,Z) \mid F=0, \partial F/\partial X=0, \partial F/\partial\alpha=0, \partial F/\partial Z=0\} = (0,1,0), \tag{20}$$

here the $\boldsymbol{B}_3$ is degenerate singular point with higher order codimension bifurcation.

As shown in Fig. 6, it is fund that the double well potentials can be obtained by varying parameters $\alpha$=0, 0.5, 1.0 and 1.5. To examine the influence of the parameter $\alpha$ on dynamics of equilibrium bifurcation of the system (13), for equilibrium bifurcation sets $\boldsymbol{B}_1$ and $\boldsymbol{B}_2$ we plotted super-critical pitch fork bifurcation diagram curve in (0, $\alpha$, $Z$) plane at bifurcation point $\boldsymbol{B}_4$ the stable branch $Z=0$ bifurcates into two stable branches at $Z=\pm(1-1/(X^2+1)^{0.5})^{0.5}$, the stable $Z=0$ now unstable, exhibiting the hyperbolic structure and sub-critical pitch fork bifurcation diagram curve in ($X$, $\alpha$, 0) plane at bifurcation point $\boldsymbol{B}_4$ he stable branch $X=0$ bifurcates into two stable branches at $X=\pm(1/(1-Z^2)-\alpha^2)^{0.5}$, the stable $X=0$ now unstable, respectively, depicted in Fig. 6(b). It should be pointed out that the singular equilibrium point $\boldsymbol{B}_4$ exhibits the higher-order multiparameter co-dimension bifurcation behavior.

### 3.2 The potential and Hamiltonia function

Integrating the nonlinear restoring force (13) over the [$X_1$, $X$], the non-dimensional potential function $PEN$ is given as

$$PEN = \int_{X_1}^{X} F_Y dx = ((X^2+\alpha^2)^{0.5}-1)^2, \tag{21}$$

where nondimensional potential $PEN$ is greater than or equal to zero dependent of nondimensional displace $X$ and geometry ratio $\alpha$.

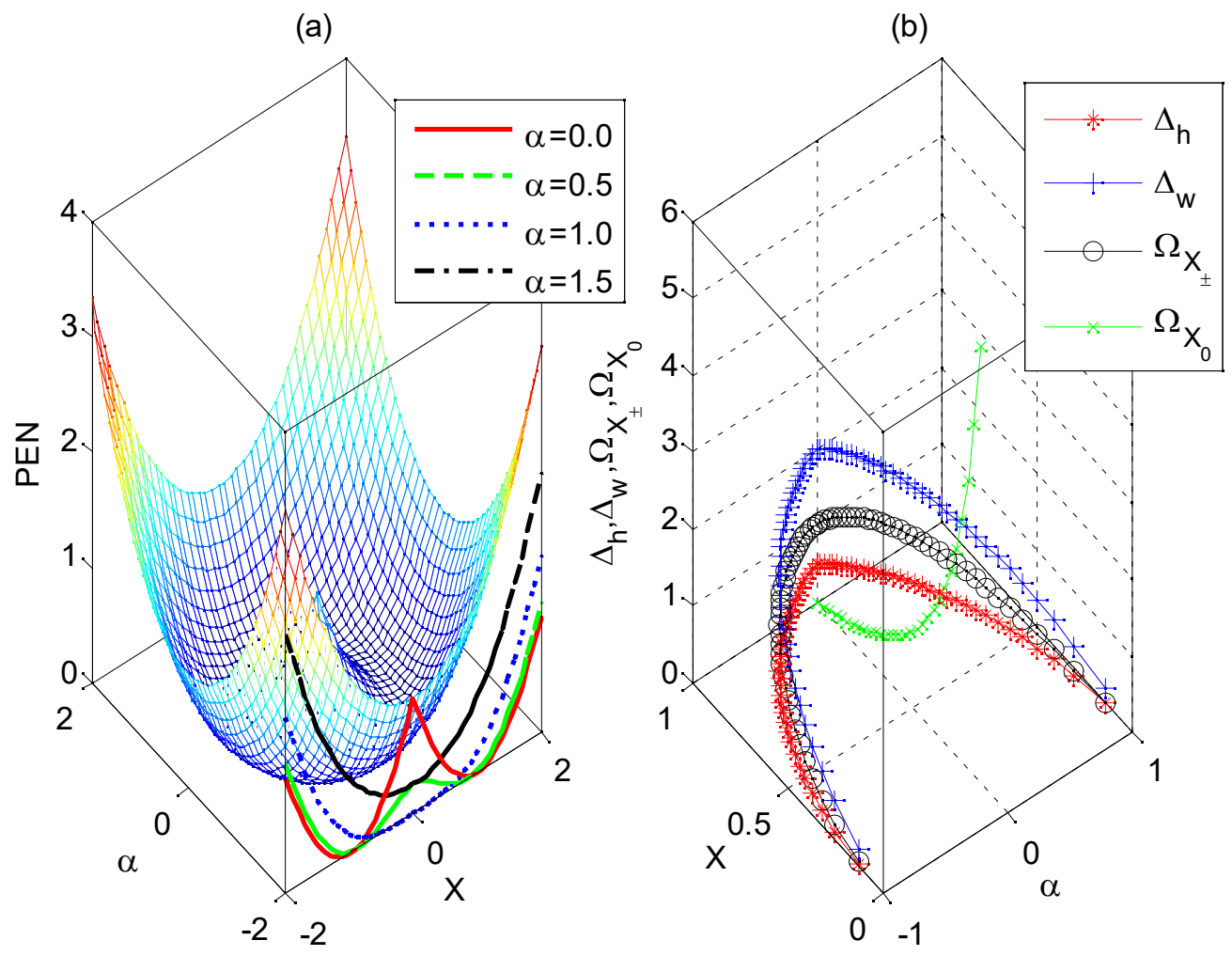

Fig. 7 Potential energy and potential barrier. (a) Non-dimensional potential-$\alpha$-displacement surfaces for Eq. (21) in ($X$, $\alpha$, $PEN$) space and the projecting potential displacement curves on ($X$,−2, $PEN$) plane with $\alpha$ = 0 (red solid curve), $\alpha$ = 0.5 (dashed curve), $\alpha$ = 1 (dotted curve), $\alpha$ = 1.5 (black dash-dotted curve) respectively. (b) Potential barriers ($\Delta_h$ and $\Delta_w$) and vibration frequecy ($\Omega_{X0}$ and $\Omega_{X\pm}$).

As presented in Fig. 7, the potential energy surface in three dimensional parameter ($X$, $\alpha$, $PEN$) space for different value of parameter $\alpha$. We see that this potential surface is similar to the shape soup-bowl-like with bulge in the middle. It is worth noticing that the multi-well behavior is similar to Duffing system dynamic as depicted on the projection plane ($X$,−2, $PEN$) in Fig. 7. We recall the the nonlinear restoring function $F_Y$ derive from the potential energy $PEN$ relies on the deformation of two oblique springs. Then it is know from the Lejeune-Dirichlet theorem that the local minimum of potential function $PEN$ corresponding to the stable equilibrium.

The solid and dashed curves show double minima located at $X_\pm=\pm(1-\alpha^2)^{0.5}$ of bistable double-wellwith vibration frequency $\Omega_0^2=\partial^2U/\partial X^2|_{X=X_\pm}=1-\alpha^2$ for $\alpha=0$ ($X_\pm=\pm1$) and $\alpha=0.5$ ($X_\pm=\pm3^{0.5}/2$), respectively. The potential barrier height $\Delta_h = (1-\alpha)^2/2$ decreased with increase geometric parameter α. The potential barrier width $\Delta_w = 2(1-\alpha^2)^{0.5}$ decreased with increase geometric parameter $\alpha$. The dashed and dashed-dotted curves show one minimum located at $X$=0 of monostable single-well with vibration frequency $\Omega_0^2 = \partial^2 U/\partial X^2 |_{X=0} = (\alpha-1)/\alpha$ for $\alpha = 1.0$ $(\Omega_0^2 = 1/3)$ and $\alpha = 1.5$ $(\Omega_0^2 = 1/3)$, respectively.

Multiplying both sides of the second equation of Eq. (13) by the first one, and integrating over [0, $X$], the Hamiltonian function for trigonal governor system (13) can be obtained as

$$H = Y^2 + \left((X^2+\alpha^2)^{0.5} - 1\right)^2 + Z^2X^2. \tag{22}$$

With help of the Hamiltonian function (22), the trajectory can be classified and analyzed for both the single stable and double stable case, as shown in Fig. 8.

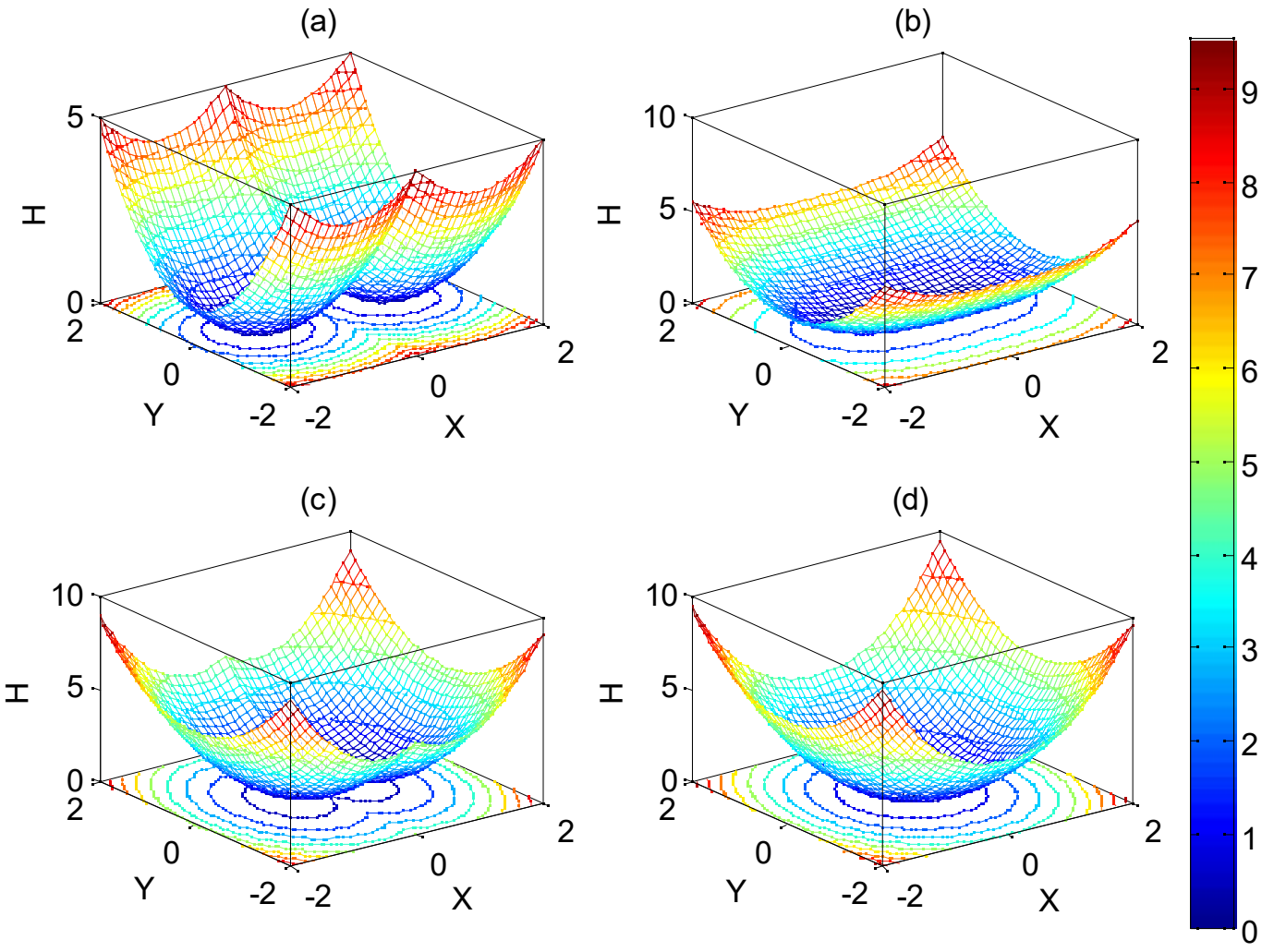

Fig. 8 Hamiltonian surfaces of $H$-velocity-diampalcement in the ($X$, $Y$, $H$) space and the contour projection on the coordinate ($X$, $Y$, 0) plane. (a)$\alpha$=0, Z=0 (b)$\alpha$=1, Z=0, (c)$\alpha$=0, Z=1, (d)$\alpha$=1, Z=1

In Fig. 8, the effect on Hamiltonian function (22) for four cases of the tuning parameters and nonlinearity strength geometry parameter $\alpha$ and non-dimensional angular velocity $Z$. Fig. 8(a) and (b) show the symmetry bistable potential well when $\alpha$=0 with nonsmooth behavior $\alpha$=0 with on the section $X$=0 and monotable well when $\alpha$=1with smooth dynamic characteristic for the angular velocity $Z$=0. Figs. 8(c) and (d) give the double well potential when $\alpha$=0 with smooth dynamic and the single well when $\alpha$=1. The phase portraits with bistable families of the circles centered at $\boldsymbol{E}_{01}$, $\boldsymbol{E}_{03}$ and homoclinic orbits are shown on the projection ($X$, $Y$, 0) planes of the Figs. 8 (a) and (c), respectively. In the same way, the Figs 8. (b) and (d) given the monostable oscillation and circle trajectories centered at $\boldsymbol{E}_{02}$ on the bottom coordinate ($X$, $Y$, 0) planes.

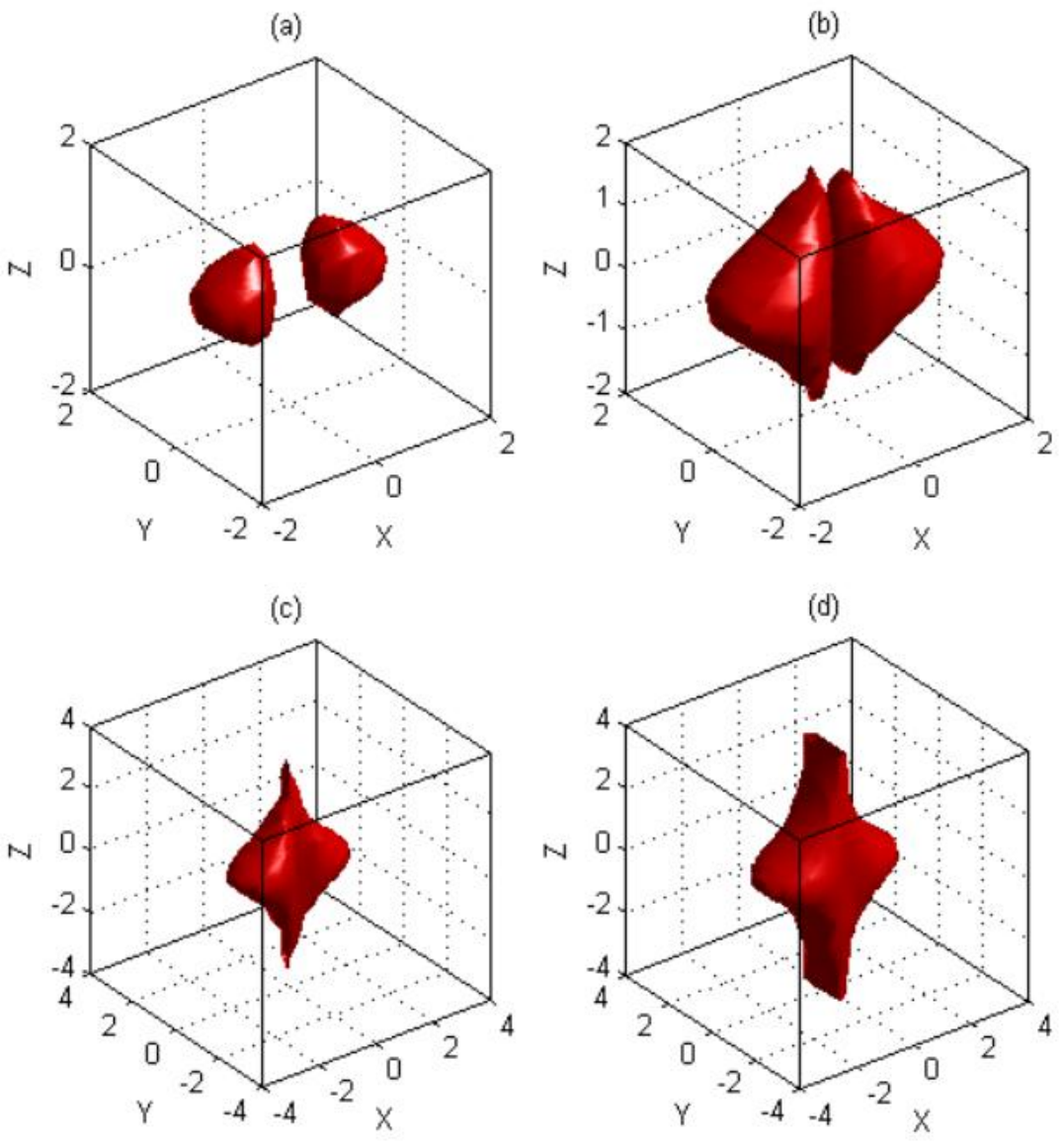

Fig. 9 Isoenergtic surfaces for different Hamiltonian value (22) with $\alpha$=0.5. (a) $H$ = 0, (b) $H$ = 0.8, (c) $H$ = 0.94, (d) $H$ = 1.7.

As shown in Fig. 9, the surfaces of phase portrait plotted via the Hamiltonian funtion Eq.(8b), at which the Hamiltonian energy is constant. Fig. 9(a) shows two chestnuts-like shapes for the half spaces $X$<0 and $X$<0 respectively. It is found that for choosing the small value of $Z$, which corresponding to the phase portraits are two circles. Two phase surfaces joined together shown Fig. 9(b), it is worth noticing that there are a separatrices of homoclinic loop on the plane of ($X$, $Y$, 0). Figs. 9(c) and (d) plotted the interwell oscillation phase surface for the periodic response.

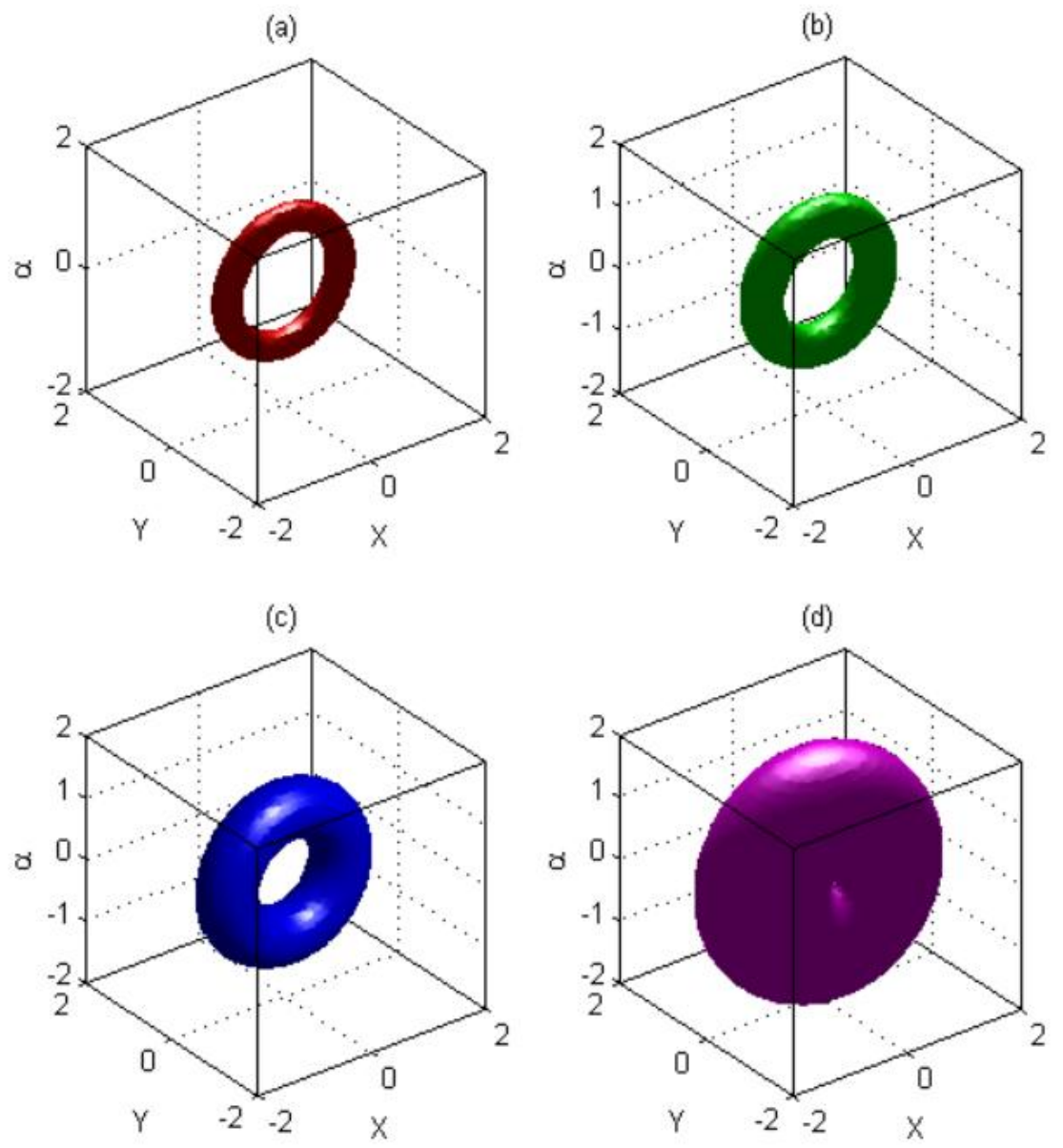

Fig. 10 Hamilton energy surface for $Z$=0 in ($X$, $Y$, $\alpha$) space  for different Hamiltonian value (22). (a) $H$=0, (b) $H$=0.1, (c) $H$=0.2, (d) $H$=0.3.

As shown in Fig. 10, the Hamilton isoenergeitc surface are plotted for geometry parameter $\alpha$ versus the non- dimensional displacement $X$ and velocity $Y$ for different non-dimensional Hamiltonian value $H$. It is most interesting that the tori of the energy surfaces centered at (0, 0, 0) for constant value of $H$. The trajectory which follow along the torus may be difficult to plot. The tori surfaces are parameterized by two angles of the longitudinal and latitudinal angles. Motion along the tori is a characteristic of multifrequency response [31].

### 3.3 The equilibrium stability of autonomous system

For the autonomous centrifugal governor system (12), letting $X'$=$Y'$ =$Z'$ =0, we obtain the equilibria for the system (12) as follows:

$$\begin{cases} \boldsymbol{E}_1=(E_{1X},E_{1Y},E_{1Z})=(X_1,Y_1,Z_1)=\left(+\delta/\gamma,0,+\left(1-1\Big/\left((\delta/\gamma)^2+\alpha^2\right)^{0.5}\right)^{0.5}\right), \\ \boldsymbol{E}_2=(E_{2X},E_{2Y},E_{2Z})=(X_2,Y_2,Z_2)=\left(+\delta/\gamma,0,-\left(1-1\Big/\left((\delta/\gamma)^2+\alpha^2\right)^{0.5}\right)^{0.5}\right), \\ \boldsymbol{E}_3=(E_{3X},E_{3Y},E_{3Z})=(X_3,Y_3,Z_3)=\left(-\delta/\gamma,0,+\left(1-1\Big/\left((\delta/\gamma)^2+\alpha^2\right)^{0.5}\right)^{0.5}\right), \\ \boldsymbol{E}_4=(E_{4X},E_{4Y},E_{1Z})=(X_4,Y_4,Z_4)=\left(-\delta/\gamma,0,-\left(1-1\Big/\left((\delta/\gamma)^2+\alpha^2\right)^{0.5}\right)^{0.5}\right), \end{cases} \tag{23}$$

where the equilibria $\boldsymbol{E}_i$ (i=1, 2, 3, 4) is fixed points of the center with the nonhyperbolic characteristic.

As shown in Fig. 11, the equilibrium surfaces are shown to reveal the parameter dependency relationships of the equilibrum $E_{iX}$ and $E_{iZ}$, the geometrical dimensionless parameters $\alpha$ and $\delta/\gamma$. From Figs. 11(a) and (c), it can be seen that the plane indicating that the equilibrium point $E_{iX}$ are linearly dependent on the parameters $\delta/\gamma$ and has nothing to do with parameter $\alpha$. Figs. 11(b) and (d) present that the equilibria $E_{iZ}$ are semicircle relationship dependent on the parameters $\delta/\gamma$ and $\alpha$.

In Fig. 12, the system exhibits the pitchfork bifurcation from the singularity point view. This bifurcation possess the basic property that as $X$ cross the value $X$=1, the number of the equilibrium solution jumps from one to two branches for $\alpha$=0 shown in Fig. 12(a). For Fig. 12(b), the system (12) has the bifurcation point at $X$=0, at which one equilibrium split into two equlibira.

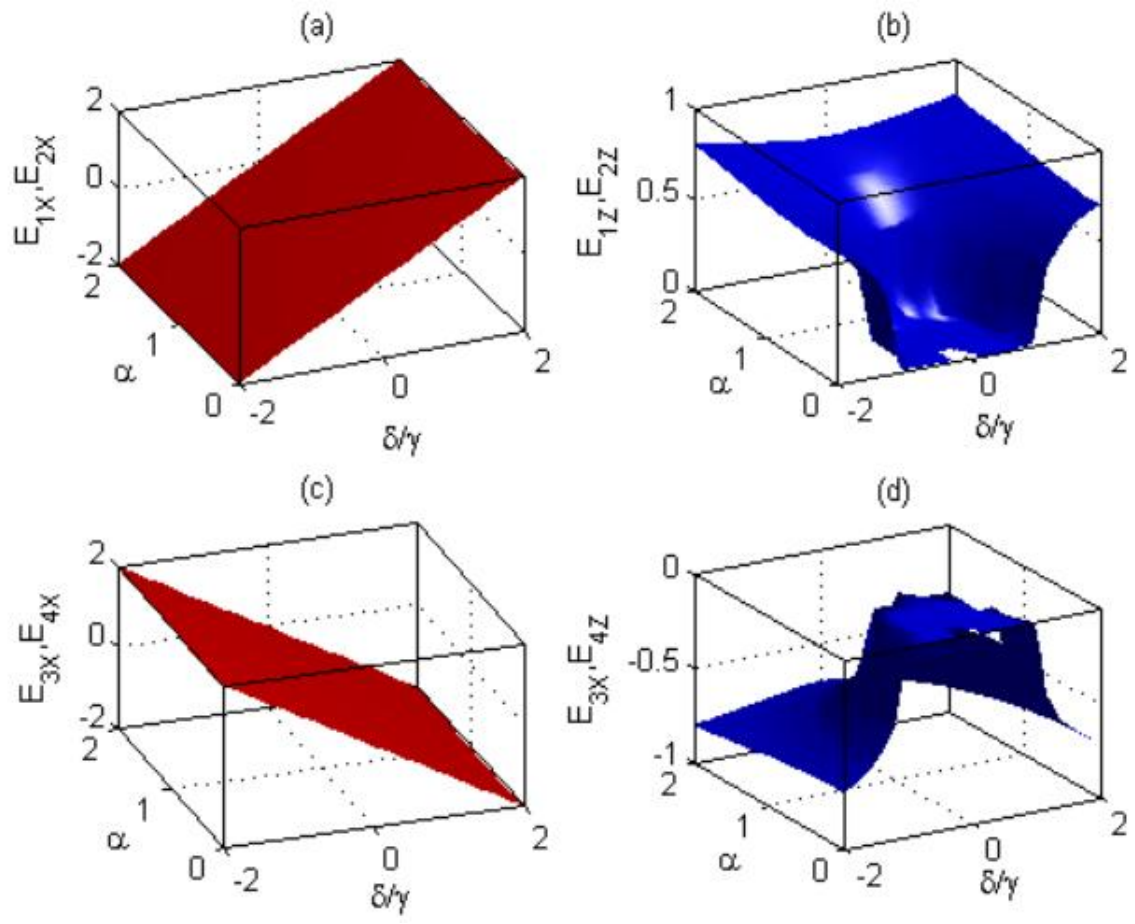

Fig. 11. Equilibrium surfaces $E_{iX}$ and $E_{iZ}$ ($i$=1, 2, 3, 4) by the Eq. (23) for the autonomous system (12) in the ($E$, $\alpha$, $\delta/\gamma$) parameter place. (a) and (c) for displace equilibrium $E_{iX}$, (b) and (d) for angular velocity equilibrium $E_{iZ}$.

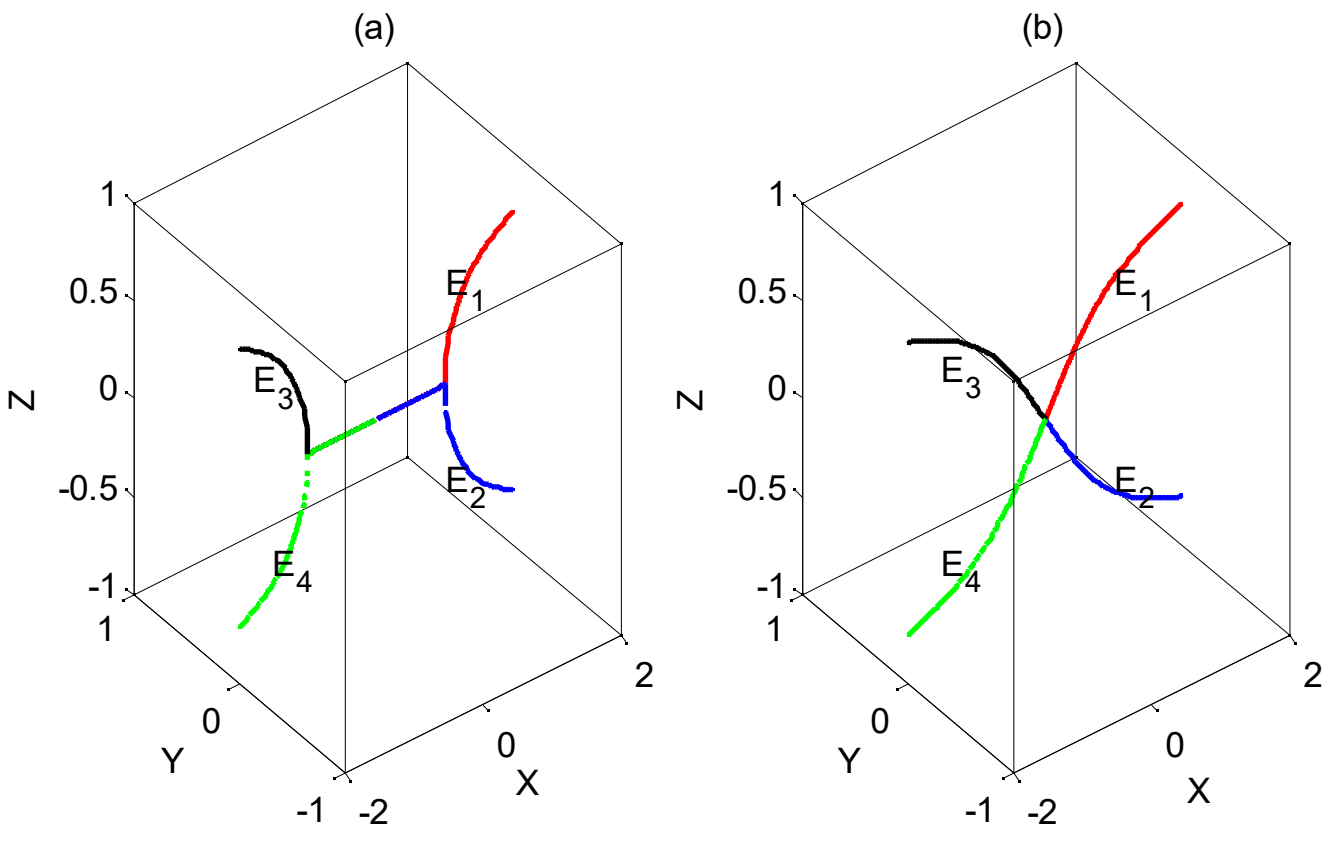

Fig. 12. Equilibrium points surface $E_i$ ($i$=1, 2, 3, 4) by the Eqs. (23) for the autonomous system (12) in the ($X$, $Y$, $Z$) place. (a) $\alpha$ = 0, (b) $\alpha$ = 1.

The Jacobian matrix of system (12) at equilibria $\boldsymbol{E}_i$ have the form

$$\boldsymbol{J}_{\boldsymbol{E}_i} = \begin{pmatrix} 0 & 1 & 0 \\ a_i & -\xi_1 & b_i \\ c_i & 0 & -\xi_2 \end{pmatrix}, \tag{24}$$

where $a_i = -1 + \alpha^2 / (X_i^2 + \alpha^2)^{3/2} + Z_i, b_i = 2Z_i X_i, c_i = -\gamma$, $i$ = 1, 2, 3, 4.

The eigenvalues of the matrix $\boldsymbol{J}_{\boldsymbol{E}_i}$T are computed to analysis stability of the equilibria $\boldsymbol{E}_i$ ($i$ = 1, 2, 3, 4) and the characteristic equation $p(\lambda)$ is illustrated by the following equation

$$p(\lambda) = A_0\lambda^3 + A_1\lambda^2 + A_2\lambda + A_3. \tag{25}$$

Then we make use of the Routh-Hurwitz criterion to determine the system (12) is stable or not. Construct the following matrix for (25) is

$$\boldsymbol{D} = \begin{pmatrix} A_1 & A_0 & 0 \\ A_3 & A_2 & A_1 \\ 0 & 0 & A_3 \end{pmatrix}, \tag{26}$$

where $A_0 = 1, A_1 = \xi_1 + \xi_2, A_2 = \xi_1\xi_2 - a_i, A_3 = a_i\xi_2 - b_i c_i$.

The parametric conditions of the root of Eq. (25) with the real parts of complex eigenvalues may be determined by follows

$$A_0 > 0, A_1 > 0, A_2 > 0, A_3 > 0, A_1A_2 - A_0A_3 > 0, \tag{27}$$

here $i = 1, 2, 3, 4$ . The parameters condition reveal that the decease the damping ratio or rotational moment of inertia, increase the mass or the torque load is no benefit for the stability of the equilibrium.

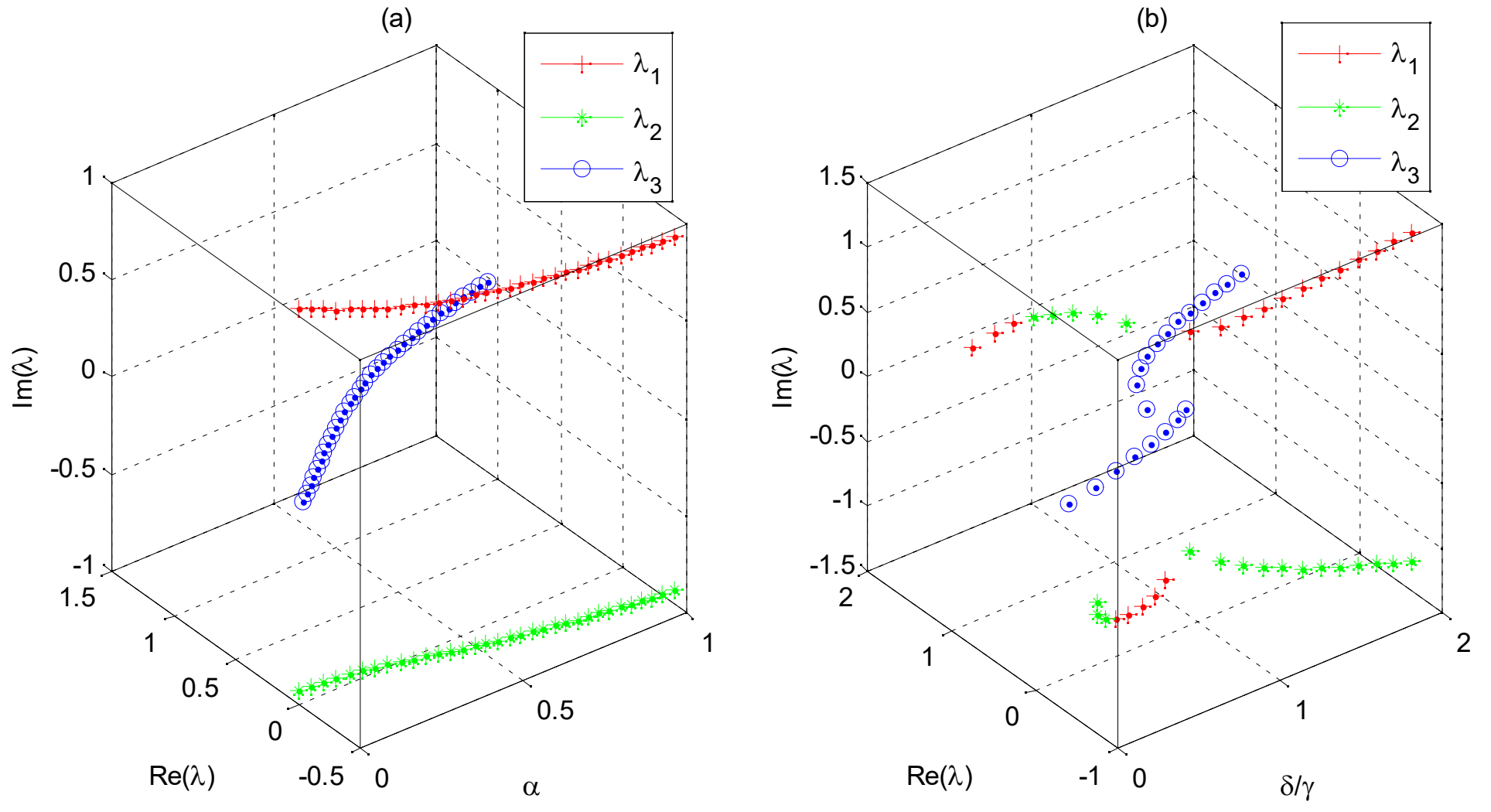

Fig. 13 Characteristic values of Jacobian matrix $\boldsymbol{J}$ of Eqs. (24). (a)$\alpha$=0; (b)$\alpha$ =1.

As shown in Fig. 13, the system (12) has the equilibria state of the saddle-focus type, i.e. $\lambda_1, \lambda_2, \lambda_3$ of the characteristic equation at $\boldsymbol{E}_i$ are such that

$$\begin{cases} \lambda_1, \lambda_2, = \rho \pm i\sigma, \sigma \neq 0, \\ \mathrm{Re}\,\lambda_3 < -\rho. \end{cases} \tag{28}$$

In this case dim $W^u_{\boldsymbol{E}_i} = 1$ and dim $W^s_{\boldsymbol{E}_i} = 2$ . Assume that the separatrices of the saddle-focus comes back to $\boldsymbol{E}_i$ as $T \to +\infty$ forming a homoclinic loop. As time $T \to +\infty$, the homoclinic trajectories tends to $\boldsymbol{E}_i$ tangentially to a two dimensional hyper place corresponding to the eigenvalues $\lambda_i (i = 1, 2)$ .

### 3.4 Numerical simulations

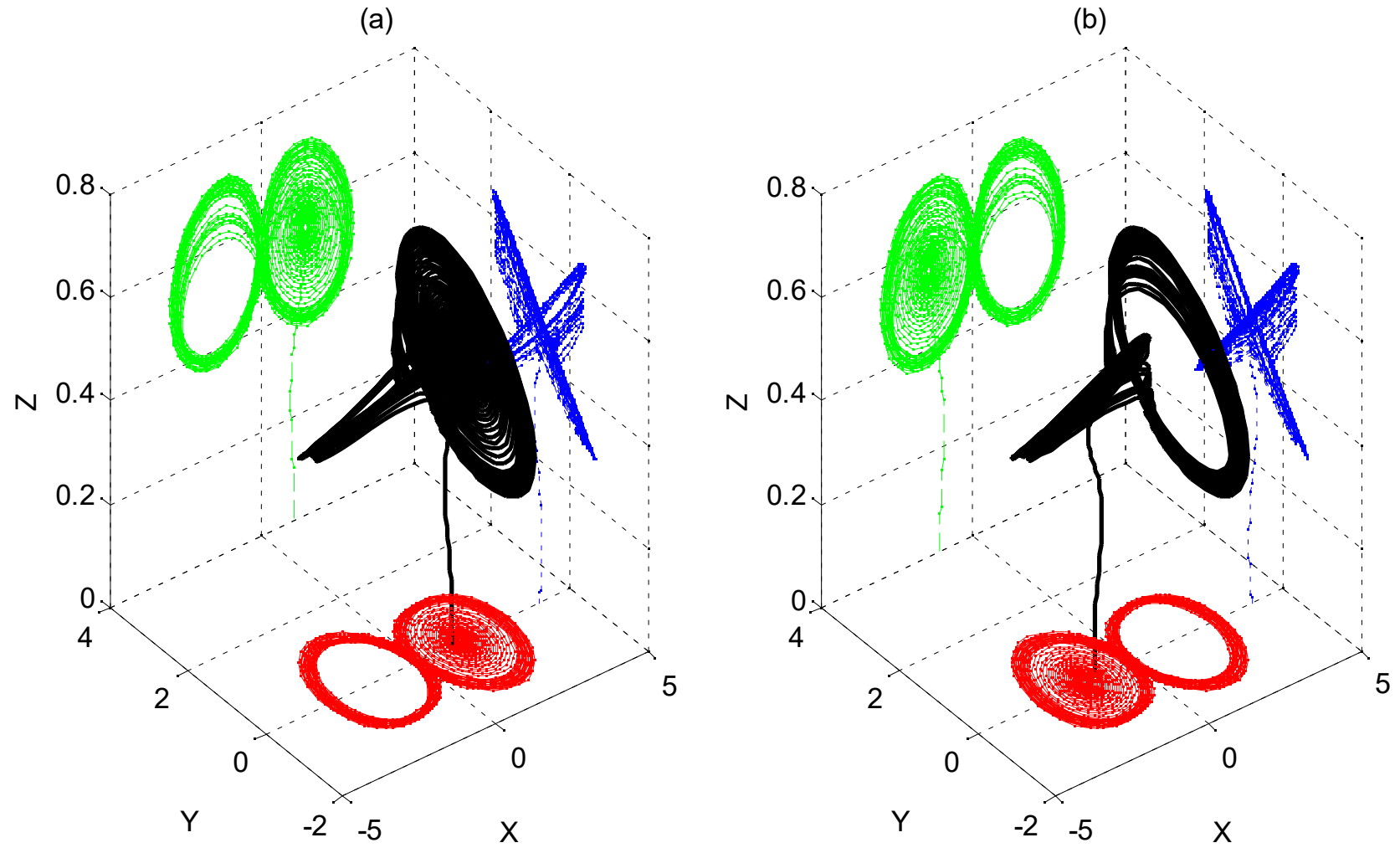

Fig. 14 Double scroll attractor of autonomous governor system (12) in the three-dimensional phase space for $\xi_1$=$\xi_2$=0.1, $\alpha$=0, $\gamma$=0.1 and $\delta$=0.2. The projections on three side planes ($X$, $Y$, 0), ($X$, 4, Z) and (4, $Y$, $Z$). (a)The trajectories for initial value (1, 0, 0). (b)The trajectories for initial value (−1, 0, 0).

We solve the equations of motion (12) numerically with a Runge-Kutta method fourth order, assuming $\xi$=0.1, $\alpha$=0, $\gamma$=0.1 and $\delta$=0.2. Fig. 14 depicts the three dimensional trajectories in ($X$, $Y$, $Z$) phase space, the cylindrical invariance tori of quasi-periodic responses are calculated for the non-autonomous system (12). It is interesting that the combination of the initial transient trajectory and an invariant torus is topologically similar to the Lorenz-like shaped structure. These two coexisting aperiodic are shown in Fig. 14(a) and (b) with bistability of double well behavior due to the trajectories on planes ($X$,$Y$) and ($X$, $Z$). In contrast, these is a butterfly-shaped attractor on the ($Y$, $Z$) projection.

To get a clearer picture of the structure of those attractors, the spatial trajectories are projected onto the the three sides of the spatial coordinate system shown in Fig.14. On the side ($X$, $Y$, 0), there are interwell oscillation with the trajectory transition from one stable to the other, which cause the large amplitude motion through the snap-through action. On the (10, $Y$, Z) projection, the trajectories from $\boldsymbol{E}_1$ and $\boldsymbol{E}_3$ almost coincide. On the ($X$, 5, Z) slice, the mass $m$ oscillates around one of the stable equilibria in the initial stage and exhibit aperiodic vibration in the terminal stage.

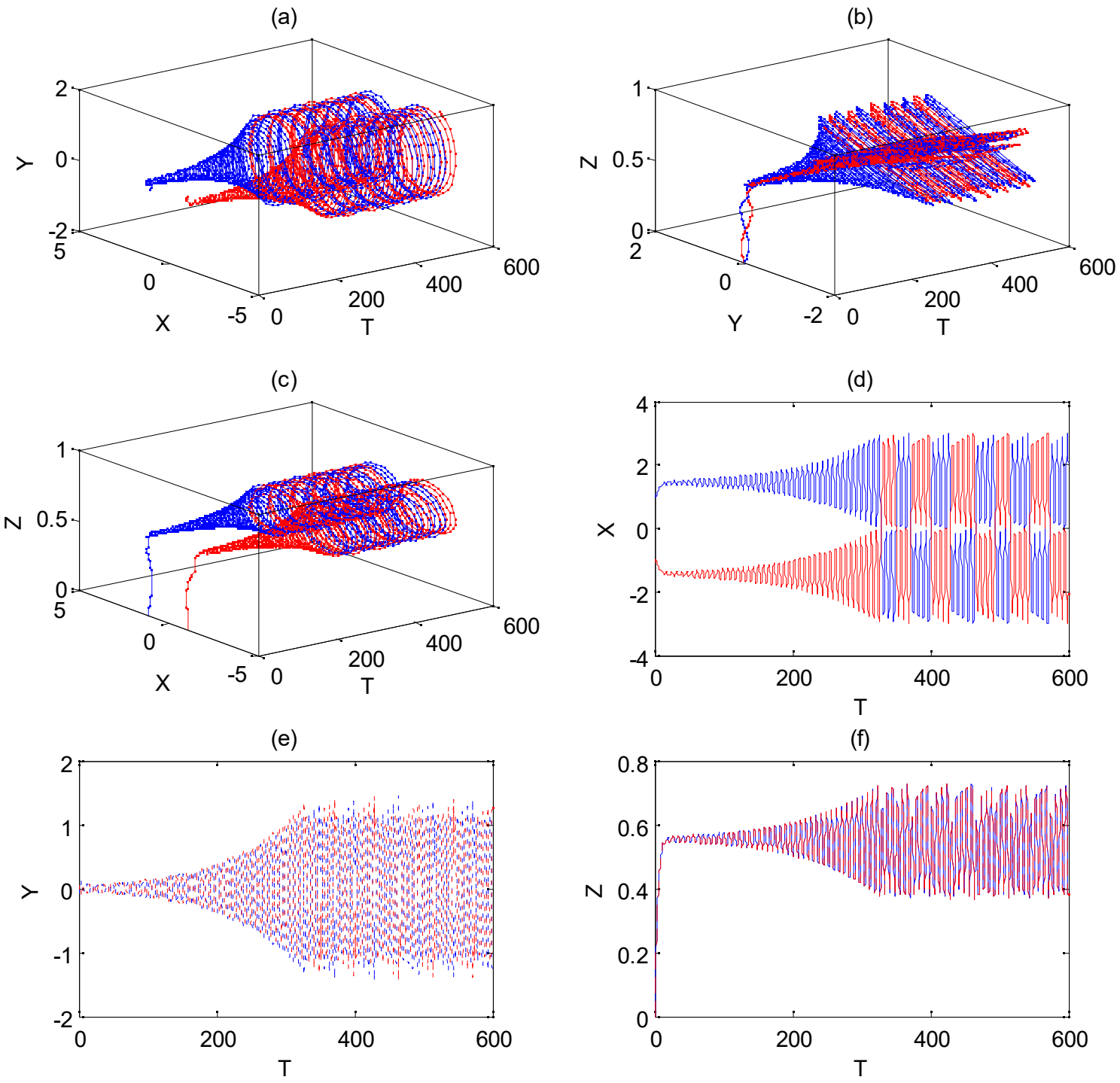

Fig. 15 The displacement-velocity-time response of the system (12) for $\xi_1=\xi_2=0.1$, $\alpha=0$, $\gamma=0.1$ and $\delta=0.2$. (a) In ($T$, $X$, $Y$ ) space, (b) In ($T$, $Y$, $Z$ ) space, (c) In ($T$, $Y$, $Z$ ) space, (d) On ($T$, $X$) plane, (e) On ($T$, $Y$) plane, (f) On ($T$, $Z$) plane

In Fig.15, The displacement-velocity-time response for the system (12) have been performed under the following system parameters for $\xi=0.1$, $\alpha=0$, $\gamma=0.1$ and $\delta=0.2$ by using numerical computation. In Fig. 15(a), it is found that the nondimensional displacement-velocity-time response increase from small to large and then to leveled off when time increasing and interlace with each other. In Figs. 15 (b) and 15(c), the rotational speed form the both equilibrium $\boldsymbol{E}_1$ and $\boldsymbol{E}_3$ firstly increase and then decrease slightly as the nondimensional time increase and finally tend to small fluctuation .

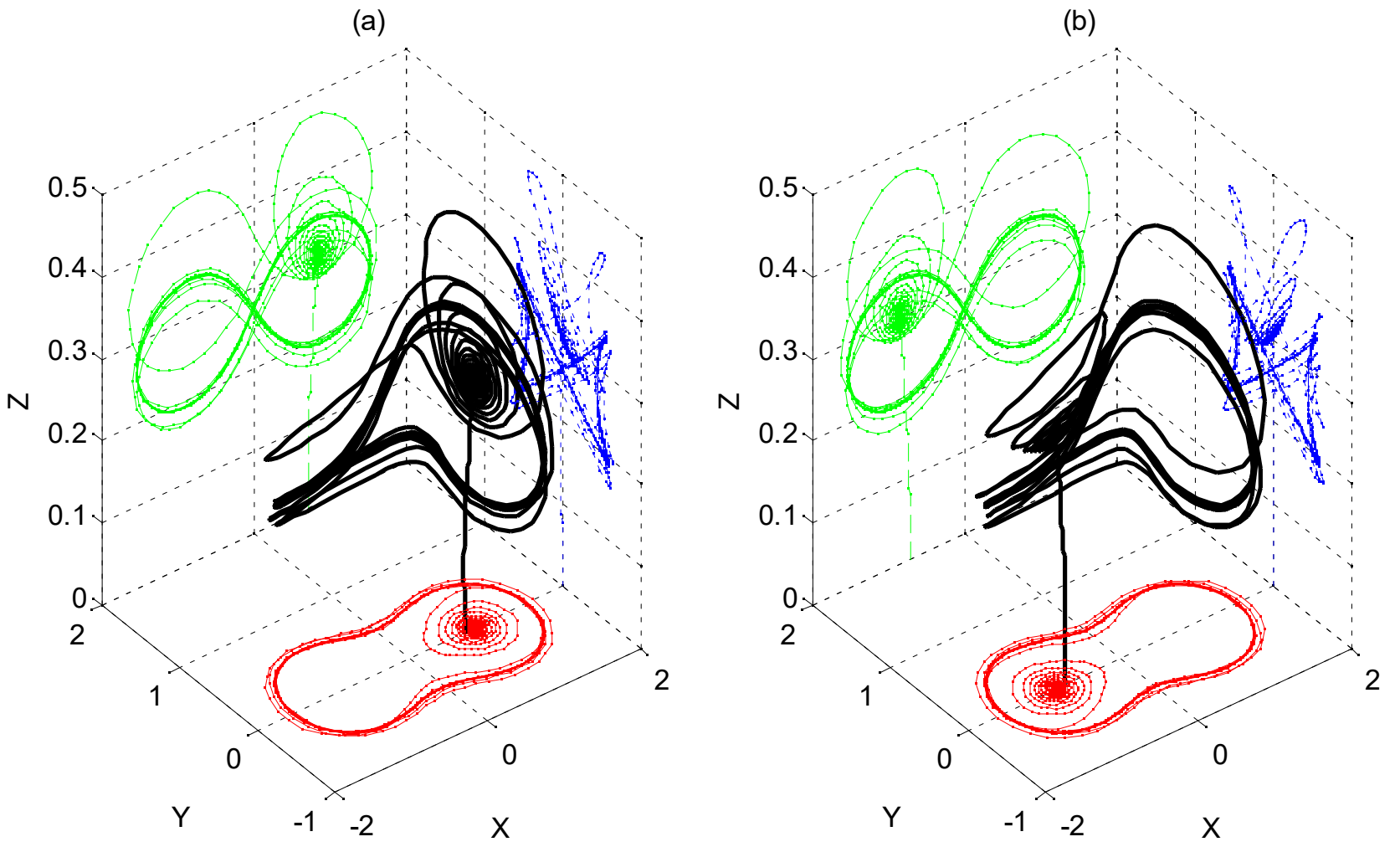

Fig. 16 The spiral periodic response for the system (12) in the three-dimensional phase portrait space for $\xi_1=\xi_2=0.05$, $\alpha=0.7$, $\gamma=0.1$ and $\delta=0.1$. Solid bold — and thin — lines denote the trajectories for equilibrium $E_1$ and $E_4$ respectively. The projections on three side planes ($X$, $Y$, 0), ($X$, 1, Z) and (10, $Y$, $Z$). (a)Initial value (0.7, 0, 0). (b)Initial value (−0.7, 0, 0).

In Fig. 16, for the parameters are fixed in $\xi$=0.05, $\alpha$=0.7, $\gamma$=0.1 and $\delta$=0.2, the three dimensional phase trajectories and projection are plotted to demonstrate the limit cycle solution. As shown in Fig.16 (a), the trajectory response is obtained for the initial condition (0.7, 0, 0) and it will be attracted to the periodic solution. In contrast for the initial condition (−0.7, 0, 0), it is found that the the phase trajectory also converges to the same closed orbit of the condition (−0.7, 0, 0). Therefor the all solutions in space converge to the same periodic response in the upper half of space of $Z$>0. In addition, the projection of phase portraits in three coordination planes are used to shown the different shape of periodic motion of autonomous system(12).

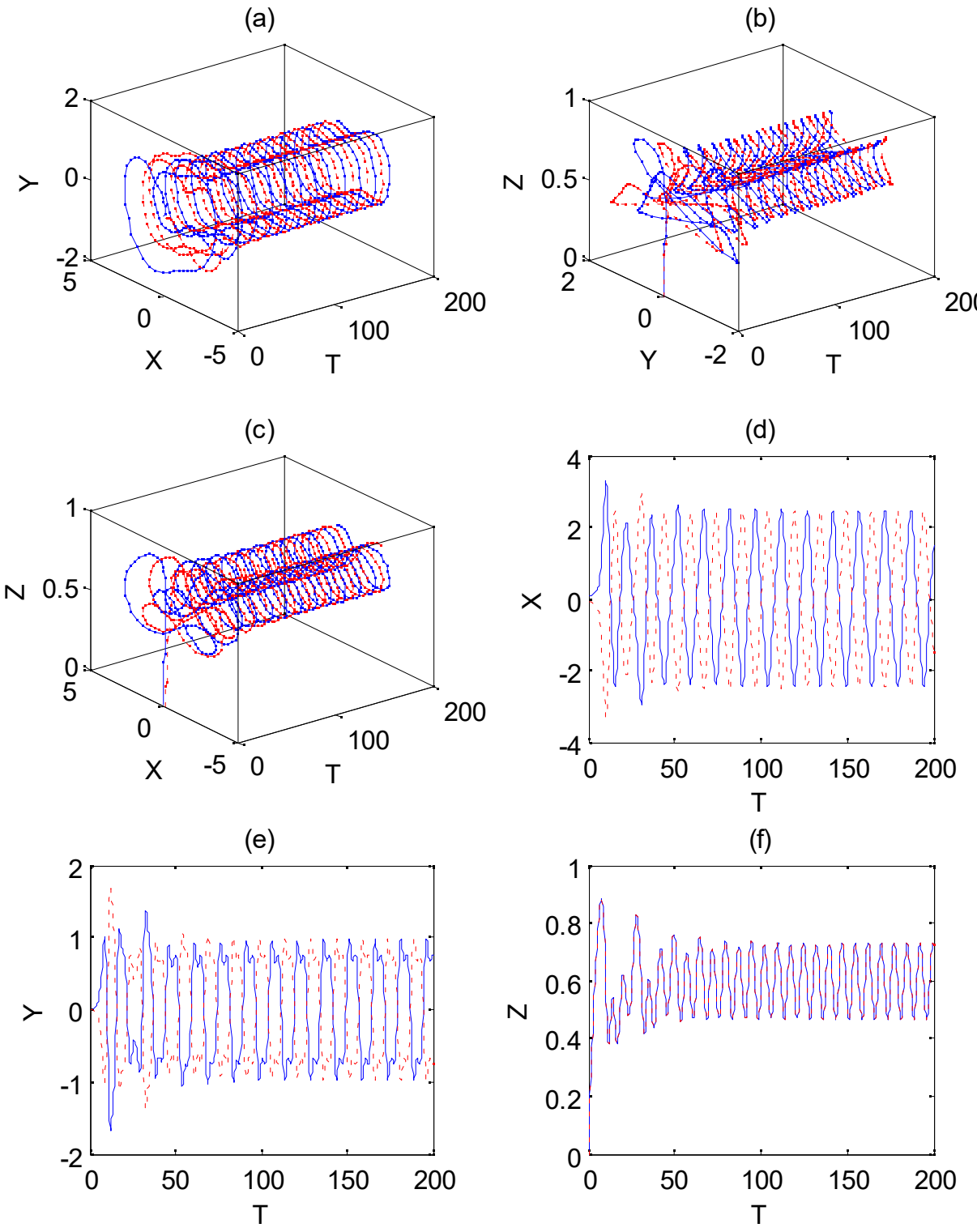

Fig. 17 The dimensionless displacement-velocity-time response of the system (12) for $\xi_1$=$\xi_2$=0.2, $\alpha$=0.7, $\gamma$=0.1 and $\delta$=0.2. (a) In spatiotemporal ($T$, $X$, $Y$ ) space, (b) In ($T$, $Y$, $Z$ ) space, (c) In ($T$, $Y$, $Z$ ) space, (d) On ($T$, $X$) plane, (e) On ($T$, $Y$) plane, (f) On ($T$, $Z$) plane

In Fig. 17, for $\alpha$=0.7, the displacement-velocity-time response and projection of the space-time ($X$, $T$) are plotted to demonstrate the periodic solution. As shown in Figs.17(a)-(c), the phase-time in three dimensional space of trajectory response are obtained for both the initial condition (0.1, 0, 0) and (−0.1, 0, 0), it is found that different conditions lead to different attraction processes to the chaotic solution. In the Figs. 17(d) and (e), it is shown that the the displacement and velocity response symmetry about the axis of time. However, Fig. 17(f) shows the angular velocity-time response is overlaps completely over the entire timeline.

## 4. The nonautonomous system

### 4.1 The criteria of chaotic threshold for three-order nonautonomous system

If the autonomous system (6) is assumed to be perturbed by periodic force $Q_1 = 2f_0 \sin(\omega_1 t)$, equation of motion of the non-autonomous trigonal centrifugal model given by

$$m\ddot{x} + c_1\dot{x} + kx(1 - b/(x^2 + a^2)^{0.5}) - m\omega^2 x = f_0 \cos\omega_1 t. \tag{29}$$

Combine the rotation Eq. (9) and the displacement Eq. (29), the non-autonomous CTG system can be written in the three dimensional form as following

$$\begin{cases} \dot{x} = y, \\ \dot{y} = -(c_1\dot{x} + kX(1 - b/(x^2 + a^2)^{0.5}))/m + \omega^2 x + f_0 \cos\omega t, \\ \dot{\omega} = (-c_2\omega - h|x| + M)/I. \end{cases} \tag{30}$$

Here we turn the autonomous system (12) into the non-autonomous system (30).

The non-autonomous CTG system (30) can be rearranged to the following three-dimensional form

$$\begin{cases} X' = Y, \\ Y' = -\xi_1 X' - X(1 - 1/(x^2 + \alpha^2)^{0.5})) + Z^2 X + F_0 \cos\Omega_1 T, \\ Z' = -\xi_2 Z - \gamma |X| + \delta. \end{cases} \tag{31}$$

where $F_0 = f_0/(kb), \Omega_1 = \omega_1/\omega_n$.

The radical nonlinear force function $F_Y$ expanded by Taylor series method, the approximately perturbed system with Duffing non-linearity is obtained as follows[30]

$$\begin{cases} X' = Y, \\ Y' = -\xi_1 X' + \eta X - \mu X^3 + Z^2 X + F_0 \cos\Omega_1 T, \\ Z' = -\xi_2 Z - \gamma |X| + \delta. \end{cases} \tag{32}$$

where $\eta = (1-\alpha)/\alpha, \mu = 1/(3\alpha^3)$.

For small parameter $F_0=\xi_1=\xi_2=0$, the homcilinic orbit is obtained as follows

$$\boldsymbol{X}^h(T) = \begin{pmatrix} X^h(T) \\ Y^h(T) \\ Z^h(T) \end{pmatrix} = \begin{pmatrix} \pm\sqrt{2/\mu}\,\text{sech}\sqrt{\lambda}T \\ \mp\sqrt{2/\mu}\lambda\text{sech}\sqrt{\lambda}T\tanh\sqrt{\lambda}T \\ -\gamma\arctan(\sinh(\sqrt{\lambda}T)) + \delta T \end{pmatrix}. \tag{33}$$

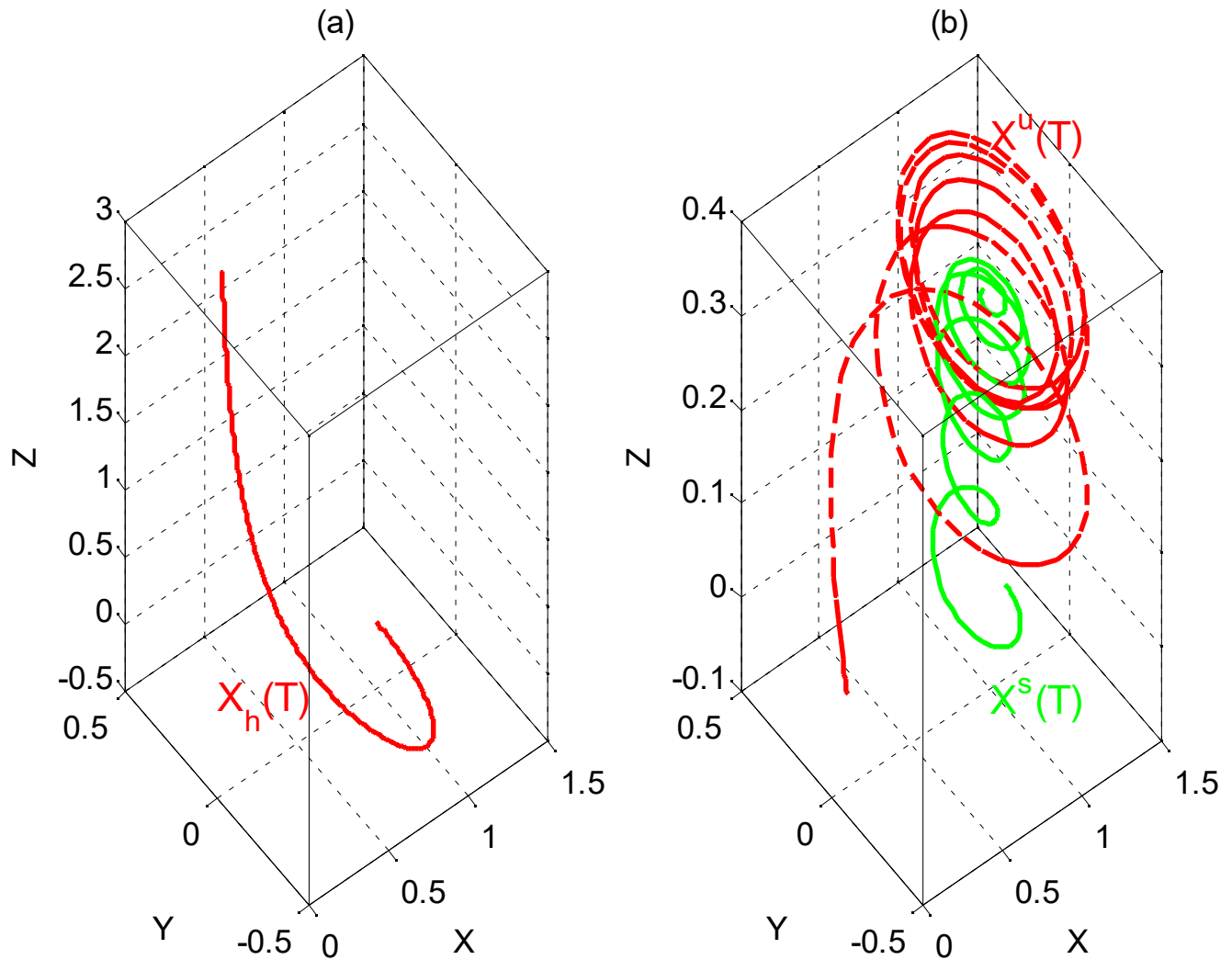

Fig. 18 Three-dimensional phase portraits in (*X*, *Y*, *Z*) space. (a)The homoclinic orbit of Eq. (33) for the autonomous system (5). (b)The trajectries for the the non-autonomous system (31).

As shown in Fig. 18, we give the homoclinic orbit for Eq. (33) for both the unperturbed case and perturbed case. See Fig. 18(a). The a spiraling curve trajectory reflects the dynamic behavior that the autonomous CTG system (12). As shown in Fig.18(b), let $\boldsymbol{X}^h$ and $\boldsymbol{X}^s$, $\boldsymbol{X}^u$ be trajectories of the autonomous and non-autonomous vector fields, respectively. We remark that $\boldsymbol{X}^s$, $\boldsymbol{X}^u$ actually depending on *T*, i.e., the perturbed vector field is non-autonomous. Therefore, the stable $\boldsymbol{X}^s$ and unstable $\boldsymbol{X}^u$ manifolds exhibit complex curves of transverse homoclinic on the projecting plane $Z=Z_0$. The three dimensional flows of the for dynamical trajectories of the perturbed CTG system (30) are plotted in Fig. 12(b).

### 4.2 Melnikov function in three-dimensional phase space

In this section, we propose a novel generalized Melnikov method, which enables us to study the Poincare map for time-periodic three-dimensional systems of the following form

$$\boldsymbol{X}' = \boldsymbol{F}(\boldsymbol{X}) + \boldsymbol{g}(\boldsymbol{X}, \boldsymbol{T}), \tag{34}$$

where $\boldsymbol{X} = \begin{pmatrix} X \\ Y \\ Z \end{pmatrix}, \boldsymbol{F}(\boldsymbol{X}) = \begin{pmatrix} Y \\ \eta X - \mu X^3 \\ 0 \end{pmatrix}, \boldsymbol{g}(\boldsymbol{X}, \boldsymbol{T}) = \begin{pmatrix} 0 \\ -\xi_1 Y + Z^2 X + F_0 \cos\Omega_1 T \\ -\xi_2 Z - \gamma |X| + \delta \end{pmatrix}.$

Equivalently, we have the follow system, letting $\boldsymbol{g}$=0, system (34) have the only one hyperbolic periodic orbit $\boldsymbol{X}_{s1}=\boldsymbol{p}_s+O(1)$. Therefore, the poincare map of the system (34) have the only one hyperbolic saddle $\boldsymbol{p}_{s1}+O(1)$. stable manifold and unstable manifold of $\boldsymbol{p}_{s1}$ do not overlap, but they are almost near the homoclinic orbit $\boldsymbol{X}^h(T-\tau)$, the trajectories in the stable and unstalble unifold are obtained

$$\begin{cases} \boldsymbol{X}^s(T,\tau)=\boldsymbol{X}^h(T-\tau)+\boldsymbol{X}_1^s(T,\tau)+O(1^1), \\ \boldsymbol{X}^u(T,\tau)=\boldsymbol{X}^h(T-\tau)+\boldsymbol{X}_1^u(T,\tau)+O(1^1). \end{cases} \tag{35}$$

At the moment $T$, the separation of between two points in stable and unstable unifolds is defined as

$$\boldsymbol{d}(T,\tau)=\boldsymbol{X}^s(T,\tau)-\boldsymbol{X}^u(T,\tau)=\boldsymbol{X}_1^s(T,\tau)-\boldsymbol{X}_1^u(T,\tau)+O(1^2). \tag{36}$$

The displacement $\boldsymbol{d}(t,\tau)$ is projected in homoclinic orbit $\boldsymbol{X}^h(T-\tau)$ of the unperturbed system, $\boldsymbol{N}$ at the $\boldsymbol{X}^h(T-\tau)$ are defined as

$$\boldsymbol{N}(T,\tau)=\boldsymbol{F}^{\perp}(\boldsymbol{X}^h(T-\tau)). \tag{37}$$

The $\boldsymbol{a}\wedge\boldsymbol{b}=(a_1\ a_2\ a_3)^{TR}\wedge(a_1\ a_2\ a_3)^{TR}$ is defined as

$$\boldsymbol{a}\wedge\boldsymbol{b}=\begin{vmatrix} \boldsymbol{i} & \boldsymbol{j} & \boldsymbol{k} \\ a_1 & a_2 & a_3 \\ b_1 & b_2 & b_3 \end{vmatrix}=(a_2b_3-b_2a_3)\boldsymbol{i}-(a_1b_3-b_1a_3)\boldsymbol{j}+(a_1b_2-b_1a_2)\boldsymbol{k}, \tag{38}$$

here $TR$ denotes matrix transpositon.

And yield to

$$\boldsymbol{d}(T,\tau)=\boldsymbol{d}\cdot\boldsymbol{N}=\boldsymbol{F}\wedge\boldsymbol{d}=\boldsymbol{d}_N^s-\boldsymbol{d}_N^u+O(1^2), \tag{39}$$

where

$$\boldsymbol{d}_N^s=\boldsymbol{F}\wedge\boldsymbol{X}_1^s,\boldsymbol{d}_N^u=\boldsymbol{F}\wedge\boldsymbol{X}_1^u. \tag{40}$$

We compute the derivative $\frac{\mathrm{d}\boldsymbol{d}_N^s}{\mathrm{d}T}$, yield to

$$\frac{\mathrm{d}\boldsymbol{d}_N^s}{\mathrm{d}T}=\frac{\mathrm{d}\boldsymbol{F}}{\mathrm{d}T}\wedge\boldsymbol{X}_1^s+\boldsymbol{F}\wedge\frac{\mathrm{d}\boldsymbol{X}_1^s}{\mathrm{d}T}. \tag{41}$$

Submitting (35) into (34), ignoring the higher order $1^2$, we obtain

$$\frac{\mathrm{d}\boldsymbol{X}_1^s}{\mathrm{d}T}=\boldsymbol{DF}\cdot\boldsymbol{X}_1^s+\boldsymbol{g}(\boldsymbol{X}_1^h(T-\tau),T) \tag{42}$$

Using (41) and (42), we obtain

$$\frac{\mathrm{d}\boldsymbol{d}_N^s}{\mathrm{d}T}=\mathrm{D}\boldsymbol{F}\cdot\boldsymbol{F}\wedge\boldsymbol{X}_1^s+\boldsymbol{F}\wedge\mathrm{D}\boldsymbol{F}\cdot\boldsymbol{X}_1^s+\boldsymbol{F}\wedge\boldsymbol{g}, \tag{43}$$

yield to

$$\frac{\mathrm{d}\boldsymbol{d}_N^s}{\mathrm{d}T}=\mathrm{tr}(\mathrm{D}\boldsymbol{F})\boldsymbol{F}\wedge\boldsymbol{X}_1^s+\boldsymbol{F}\wedge\boldsymbol{g}, \tag{44}$$

also

$$\frac{\mathrm{d}\boldsymbol{d}_N^s}{\mathrm{d}T}=\mathrm{tr}(\mathrm{D}\boldsymbol{F})\boldsymbol{d}_N^s+\boldsymbol{F}\wedge\boldsymbol{g}, \tag{45}$$

and integrating the differential equations (45) of first order linear nonhomogeneous, from $\tau$ to $+\infty$,we have

$$\boldsymbol{d}_N^s(+\infty,\tau)-\boldsymbol{d}_N^s(\tau,\tau)=\int_\tau^{+\infty}\boldsymbol{F}\wedge\boldsymbol{g}\exp\left(-\int_0^{T-\tau}\mathrm{tr}(\mathrm{D}\boldsymbol{F}(\boldsymbol{X}^h(T)))dT\right). \tag{46}$$

Since $\boldsymbol{p}_s$ is a hyperbolic saddle points, using the eqs. (38) we obtained

$$\boldsymbol{d}_N^s(+\infty,\tau)=\boldsymbol{F}(X^h(+\infty,\tau))\wedge\boldsymbol{X}_1^s=\boldsymbol{F}(\boldsymbol{p}_s)\wedge\boldsymbol{X}_1^s=0. \tag{47}$$

Then we have

$$d_N^s(\tau,\tau) = -\int_\tau^{+\infty} F \wedge g\exp\left(-\int_0^{T-\tau} \mathrm{tr}(\mathrm{D}F(X^h(T)))dT\right). \tag{48}$$

In the same way, we have

$$d_N^u(\tau,\tau) = -\int_{-\infty}^{\tau} F \wedge g\exp\left(-\int_0^{T-\tau} \mathrm{tr}(\mathrm{D}F(X^h(T)))dT\right). \tag{49}$$

Finally, we define the Melnikov function

$$M(\tau) = -\int_{-\infty}^{+\infty} F(X^h(T-\tau)) \wedge g(X^h(T-\tau),)\exp\left(-\int_0^{T-\tau} \mathrm{tr}(\mathrm{D}F(X^h(T)))dT\right)dT. \tag{50}$$

Gives

$$d_N(\tau,\tau) = -M(\tau) + O(1^2). \tag{51}$$

Applying integral variable transformation, yields

$$M(\tau) = -\int_{-\infty}^{+\infty} F(X^h(T)) \wedge g(X^h(T), T+\tau)\exp\left(-\int_0^{T-\tau} \mathrm{tr}(\mathrm{D}F(X^h(T)))dT\right)dT, \tag{52}$$

where

$$f \wedge g = \begin{vmatrix} i & j & k \\ Y & \lambda X(1-X^2) & 0 \\ 0 & -\xi_1 Y + F_0 \sin\omega_1 T & -\xi_2 Z - \gamma|X| - \delta \end{vmatrix} \tag{53}$$

$$= i\left(\lambda X(1-X^2)(-\xi_2 Z - \gamma|X| - \delta)\right) - j\left(Y(-\xi_2 Z - \gamma|X| - \delta)\right) + k\left(Y(-\xi_1 Y + F_0 \sin\omega_1 T)\right).$$

Integral equation (53), we obtain the Melnikov function [31,32] as follows

$$M(\tau) = \begin{pmatrix} M_X(\tau) \\ M_Y(\tau) \\ M_Z(\tau) \end{pmatrix} = \begin{pmatrix} 0 \\ 0 \\ -4\lambda^{3/2}/(3\lambda\mu)\xi_1 \pm \pi F\omega_1\sqrt{2/\lambda}\,\mathrm{sech}(\pi\omega_1/(2\sqrt{\eta}))\sin\omega_1\tau \end{pmatrix}, \tag{54}$$

where Melnikov's function $M(\tau)$ measure the separation and intersection of the stable and unstable manifolds on the cross-section for the CTG system (32).

For the parametrization of homoclinic trajectory $X^h(T)$, it is found that fixing $Z$ and varying $T$ corresponding to the separation of on fixed cross-section. The Melnikov function $M(\tau)$ is the measurement of the distance between $X^s$ and $X^u$. The zeros of the Melnikov function $M(\tau)$ correspond to transverse homo-clinic points of the three-dimensional map meant to chaotic motion has taken place.

Then the chaotic condition of critical threshold for non-autonomous TCG system (32) is illustrated as following form [33]

$$F_0 > \frac{4\lambda^{3/2}\xi_1}{3\pi\Omega_1\sqrt{2\lambda}}\cosh\frac{\pi\Omega_1}{2\sqrt{\eta}} = R(\xi_1,\Omega_1). \tag{55}$$

The graph of the critical curves Eq. (55) of the chaotic boundaries of Melnikovian method for system (32) obtained is shown in Fig. 19(a) for the different value of non-dimensional distance parameter $\xi = 0.1$, 0.2, 0.3, 0.4. In the area above the detected critical boundaries, chaotic motions will generated. In Fig 19(b) we plot the critical surface of parametric condition Eq. (55) for intersection of the stable and unstable manifolds. Thus Eq. (55) is the threshold condition for the chaos motion of the damped and forced TCG system (32) in parameter ($F_0$, $\xi$, $\Omega_1$) space.

In order to examine the the influence of the parameters $F_0$, $\xi_1$, $\Omega_1$ on the chaotic condition (55) from the different parameter perspective, we construct the cross sections of the critical surface for different value of damping coefficient $\xi_1$=0.1, 0.2, 0.3, 0.4, on the projection ($F_0$, 2, $\Omega_1$) parameter plane. In the same cross- sections of critical surface for $\Omega_1$= 0.1, 1,1.5, 3 on the projection ($F_0$, $\xi_1$, 0) plane. cross-sections of critical surface for $F_0$=0.1, 0.2, 0.3, 0.4 the projection (5, $\xi_1$, $\Omega_1$) parameters plane. In the area above the critical surface, chaotic motions sill be generated, i.e. in the case of $F_0 > R(\xi_1,\Omega_1)$.

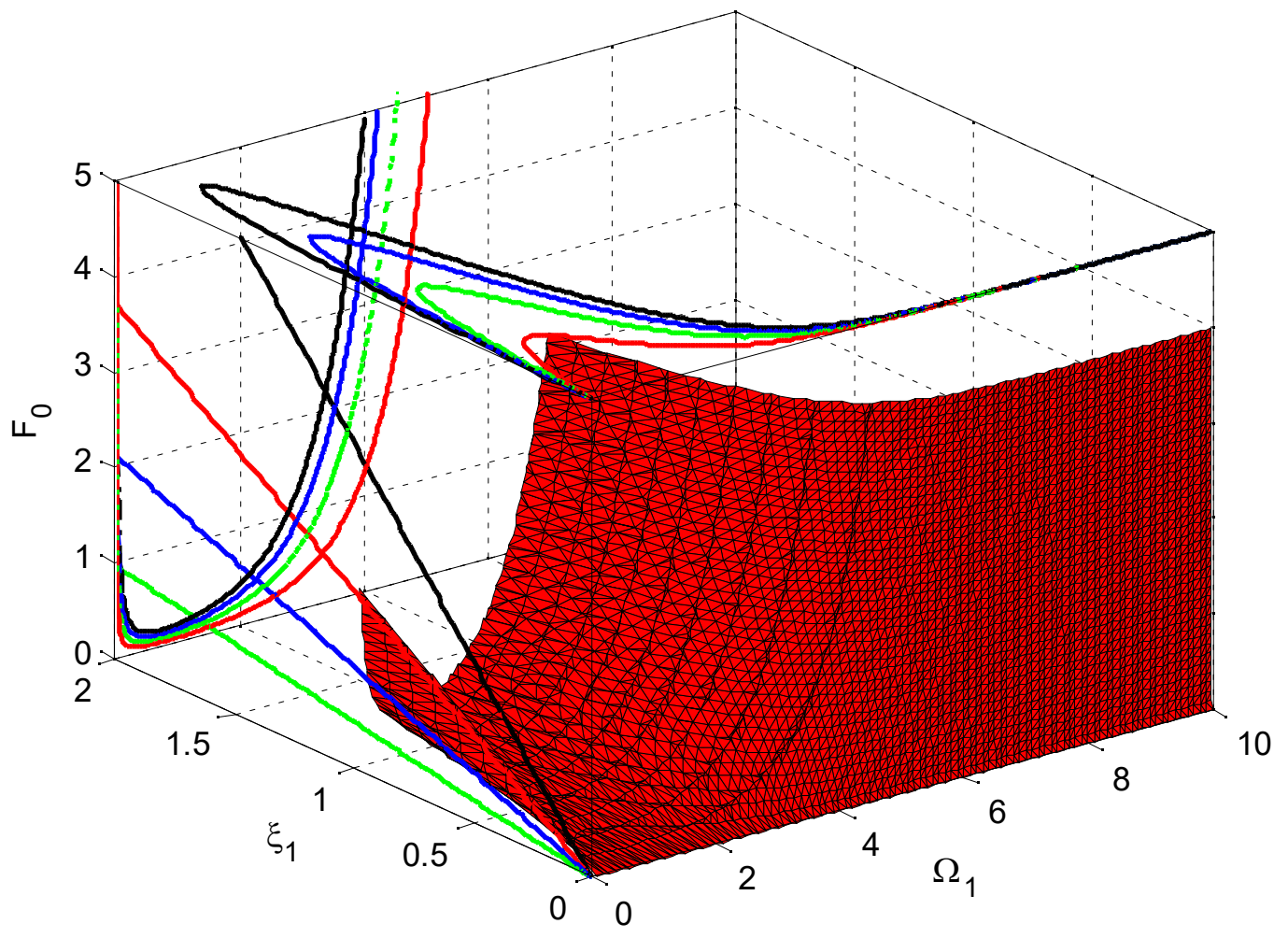

Fig. 19 Chaotic critical surface for non-autonomous system (32) with $\alpha$=0.5 in the ($F_0$, $\xi_1$, $\Omega_1$) parameters space. Cross-sections of critical surface for $\xi_1$=0.1, 0.2, 0.3, 0.4 on the projection ($F_0$, 2, $\Omega_1$) plane. cross- sections of critical surface for $\Omega_1$= 0.1, 1,1.5, 3 on the projection ($F_0$, $\xi_1$, 0) plane. The cross- sections of critical surface for $F_0$ = 0.1, 0.2, 0.3, 0.4 the projection (5, $\xi_1$, $\Omega_1$) parameters plane.

### 4.3 Chaotic motion of the non-autonomous CTG system

We solve the equation of motion (31) numerically with a Runge-Kutta method using ODE45 routine to verify the efficiency of the criteria (55). In Fig.19, as can be seen in ($X$, $Y$, $Z$) phase portraits with $F_0$ = 0.1 and the chaotic attractor are calculated for the non-autonomous system (31). The perturbed system demonstrates the regions of chaotic solution in the hyper-space of the initial system parameters.

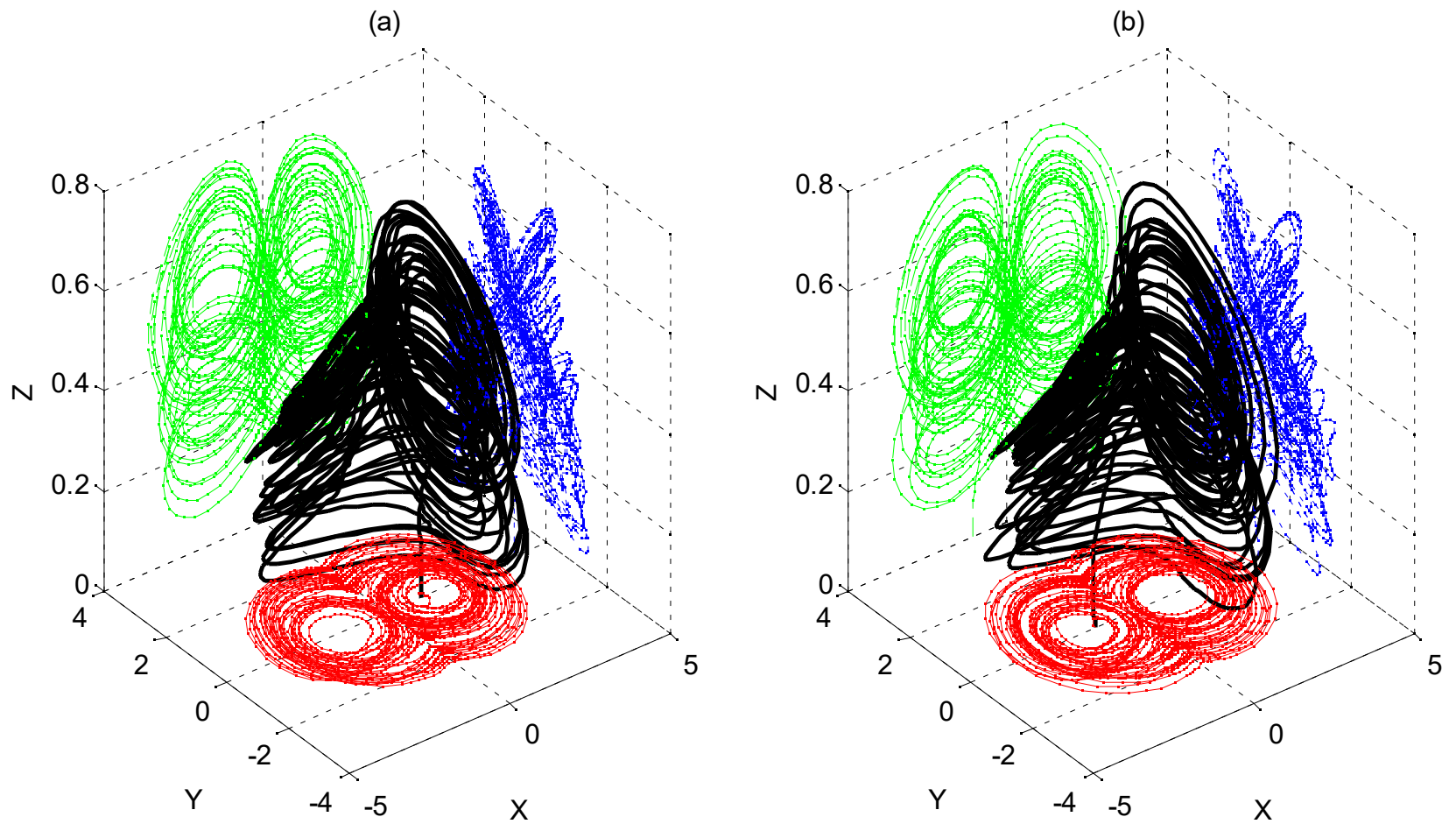

Fig. 20 Stereoscopic view of TCG attractor for the perturbed system (31) with $\xi_1$=$\xi_2$=0.1, $\alpha$=0, $\gamma$=0.1, $\delta$=0.1, $\Omega_1$=1, $F_0$=0.2. (a)Phase portraits space in the initial value(1,0, 0). (b)Poincare section in the initial value(-1,0, 0).

In Fig. 20, the chaotic response of phase portraits of the perturbed system (31) for fixed parameters of $\xi_1$=$\xi_2$=0.1, $\alpha$=0, $\gamma$=0.1, $\delta$=0.1, $\Omega_0$=1, $F_0$=0.2. It is found that the non-autonomous system (31) with bistable well exhibit aperiodic attractors with different initial condition. The phase portrait diagrams of $X$ versus $Y$, $Y$ versus $Z$ and $Y$ versus $Z$ in Figs. 20(a) and (b) show that the non-autonomous CTG system go into the chaotic vibration.

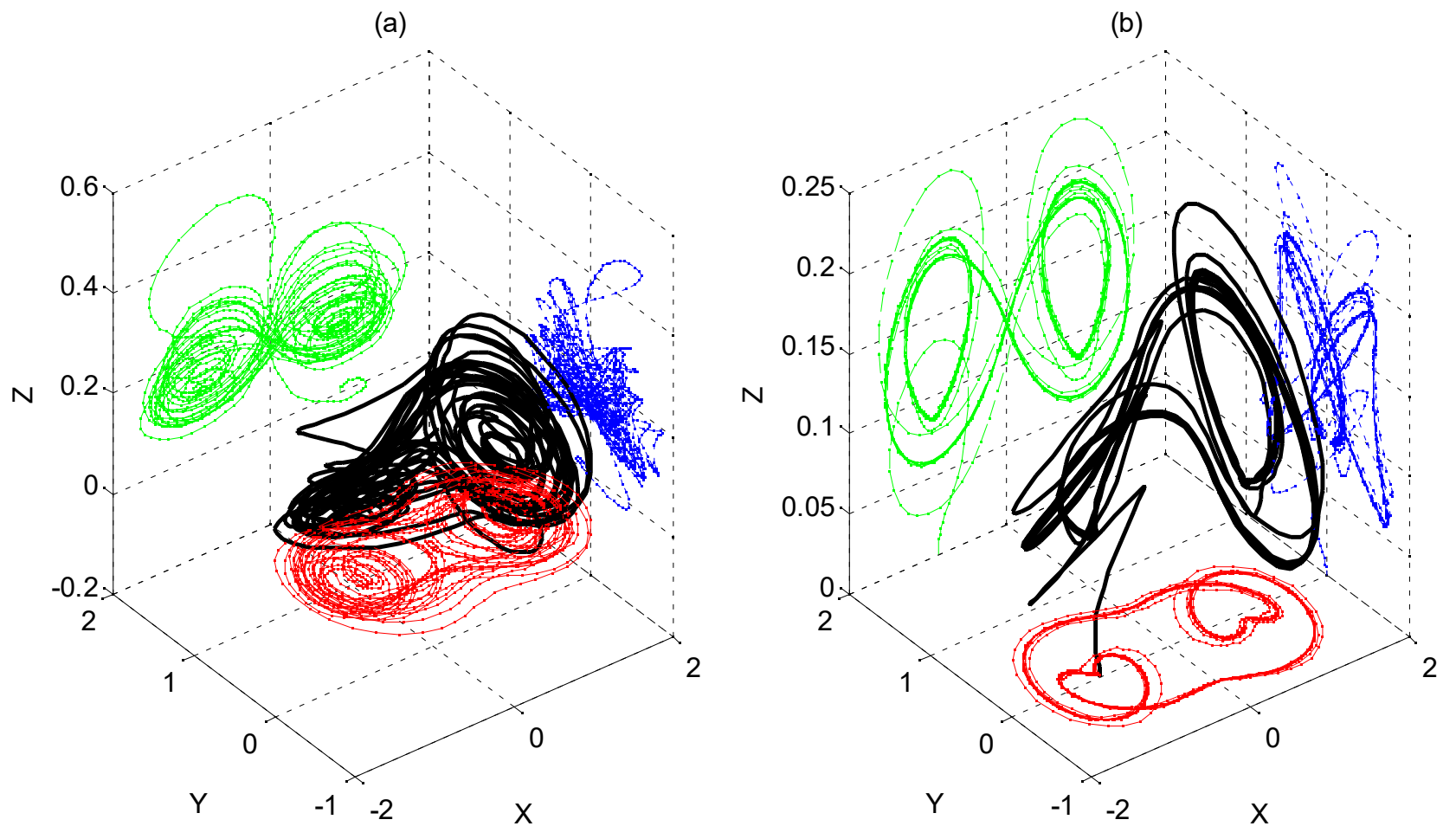

Fig. 21 The TCG attractor in the 3-d space for the perturbed system (31) with $\xi_1=\xi_2=0.1$, $\alpha=0.7$, $\gamma=0.1$, $\delta=0.2$, $\Omega_0=0.7$, $F_0=0.2$. (a)Phase portraits space in the initial value(0.1, 0, 0). (b)Poincare section in the initial value(−0.1, 0, 0).

In Fig. 21, the chaotic state of phase plane of the perturbed system (31) for fixed parameters of $\xi_1=\xi_2=0.1$, $\alpha=0.7$, $\gamma=0.1$, $\delta=0.1$, $\Omega_0=0.7$, $F_0=0.2$. It is found that the nonautomous system (31) with bi-stable well case exhibit chaotic and period doubling motion. It is found that the CTG system displays various response structures due to the different value of initial value ($X_0$, $Y_0$, $Z_0$).

## 5. Experimental study

The experimental platform of vibration test is shown in Fig. 22(a) and the prototype of the CTG system assembled manually according to the proposed design concept. The fabricated prototype is consist of air fan for providing the wind, a fan blade link with two masses, two column-helix springs connect to two masses and laser sensor.

The fan blades, the mass and the spring all rotate on the shaft, which drive come from the blowing force of the air fan. The laser Doppler vibrometer is used to measure the displacement of the mass and angular signal of the fan. The air fan have a diameter of 8mm and the fan blade with the diameter of 7.5mm. The shaft with a diameter 2mm of and length of 100mm. The two springs have the same diameter of 5mm and length of 40mm. The baffle plate with length of 80mm and width of 60mm attache to the masses to feedback the displacement of plate and control wind flow of the air fan. Fig. 22(b) give the schematic diagram of the experimental setup. The simulation of displacement, velocity and angular velocity were validated against measured date.

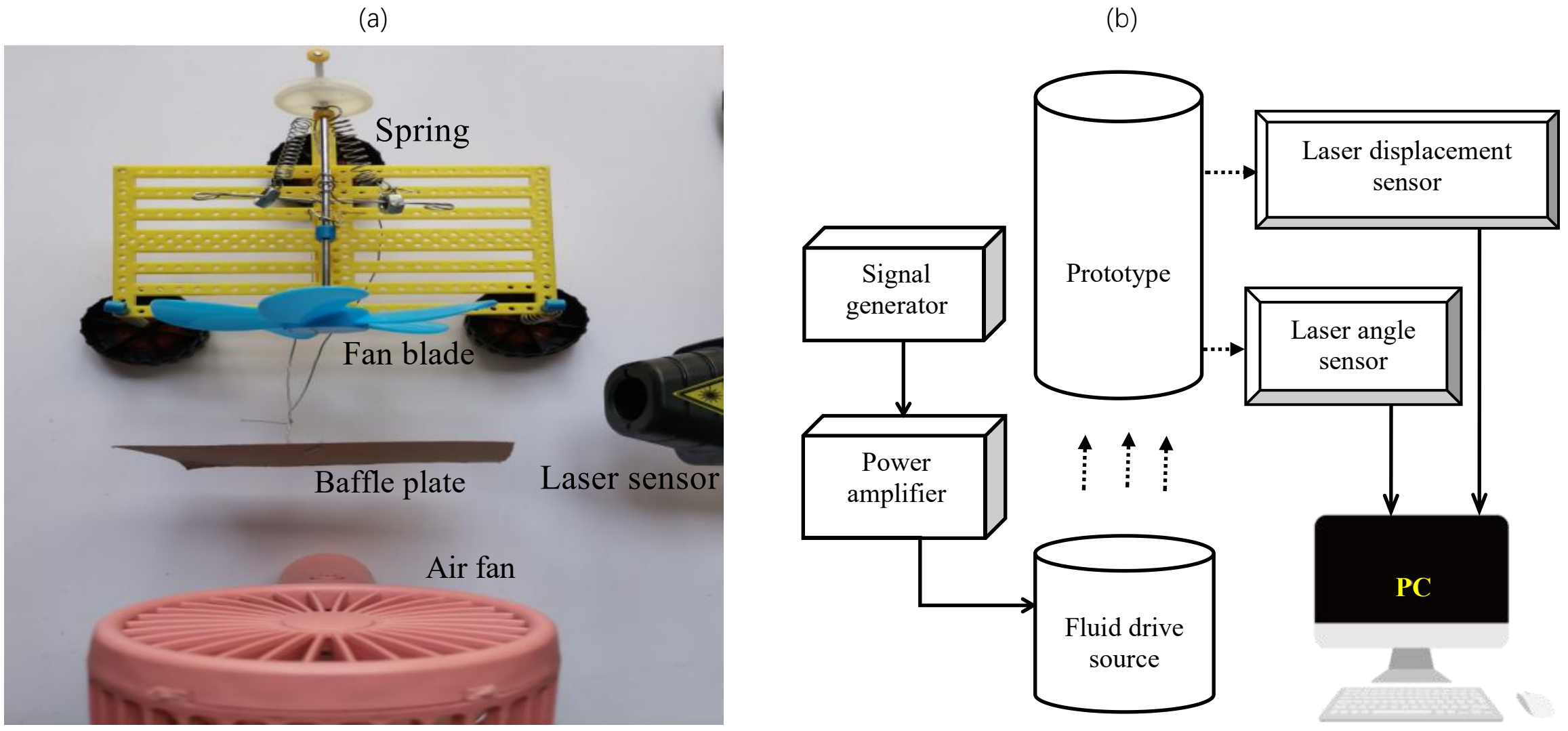

Fig 22. Experimental setup for measure displacement and rotational velocity. (a)Photo of fabricated prototype. (b)Schematic diagram.

The material and geometrical parameters of the CTG system are listed in Table 1. The experiments were conducted for various input power, which come from the air fan of the different rotational speed 850r/min, 950r/min and 1050rad/min. The results for the CTG system, the displacement, velocity and angular velocity are calculated and shown in Figs. 23-25. Fig. 23 shows the trajectories of system (10) for the initial conditions (±($b^2$−$a^2$)$^{0.5}$, 0, 0) and the steady state response for focus points (±0.026, 0, 39.949). The test results of steady state response are (±0.02, 0, 38). Fig. 24 plots the three dimensional space phase portraits of system (10) for the initial conditions (±($b^2$−$a^2$)$^{0.5}$, 0, 0) and the focus points (±0.031, 0, 49.938). The steady state response of test results are (±0.03, 0, 52). Fig. 25 shows three dimensional trajectories of system (10) for the initial condition (±($b^2$−$a^2$)$^{0.5}$, 0, 0) and the steady state response for focus points (±0.040, 0, 59.921). The test results of steady state response are (±0.035, 0, 62). It is fond that the both numerical and experimental result are in good agreement.

Table 1. Practical parameters of the CTG device

| Parameters | Symbol | Values | Unit |
|---|---|---|---|
| Mass of sliding blocks | $m$ | 0.02 | kg |
| Stiffness of spring | $k$ | 20 | N/m |
| Damping of masses and bars | $c_1$ | 20 | N·s/m |
| Damping of shaft and bearing | $c_2$ | 20 | N·s·m/rad |
| Distance of simply supports | $a$ | 0.03 | m |
| Free length of springs | $b$ | 0.02 | m |
| Moment of inertia | $I$ | 0.0002 | kg·m$^2$ |
| Load torque | $M$ | 4 | N·m |
| Force constant | $h$ | 0.2 | N |

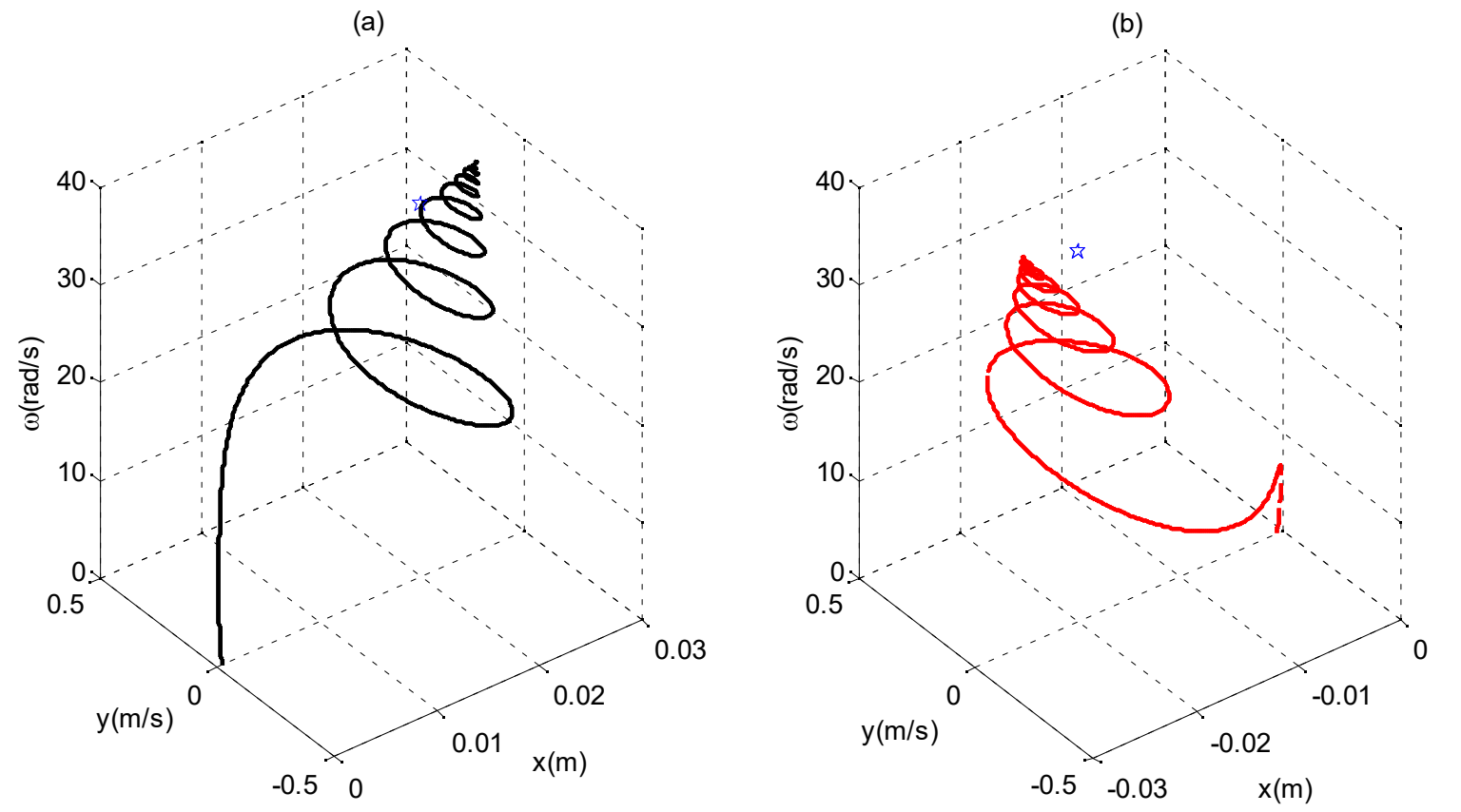

Fig 23. The experimental results from the protype system for $n$=850r/min. Solid curves and pentagram denote simulated and exprimental results respectivelly. (a)Phase potrait for inital condition of $E_1$, (b)Phasepotrait for inital condition of $E_3$.

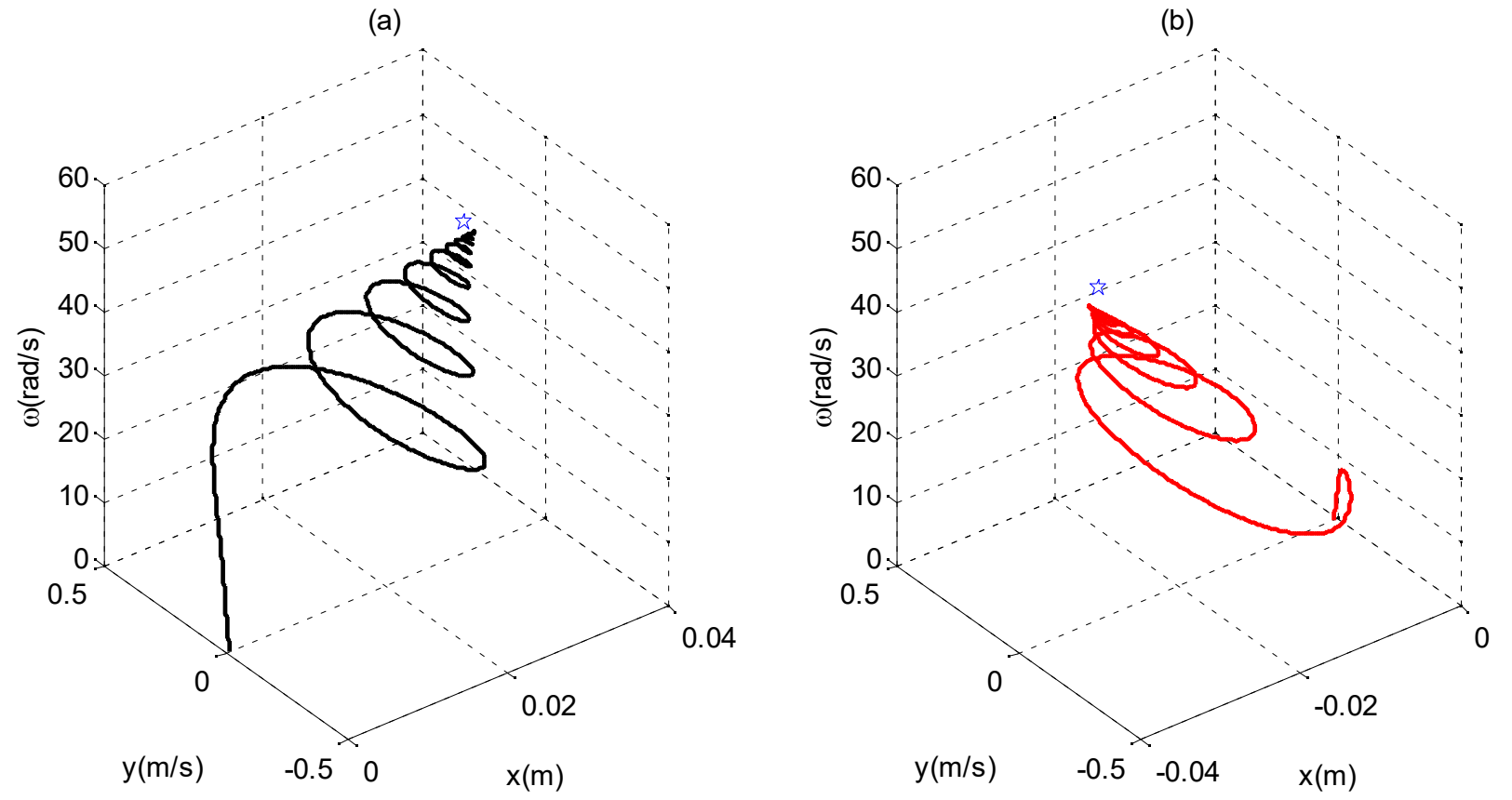

Fig 24. The experimental results from the protype system for $n$=950r/min. Solid curves and pentagram denote simulated and exprimental results respectivelly. (a)Phasepotrait for inital condition of $E_1$, (b)Phasepotrait for inital condition of $E_3$.

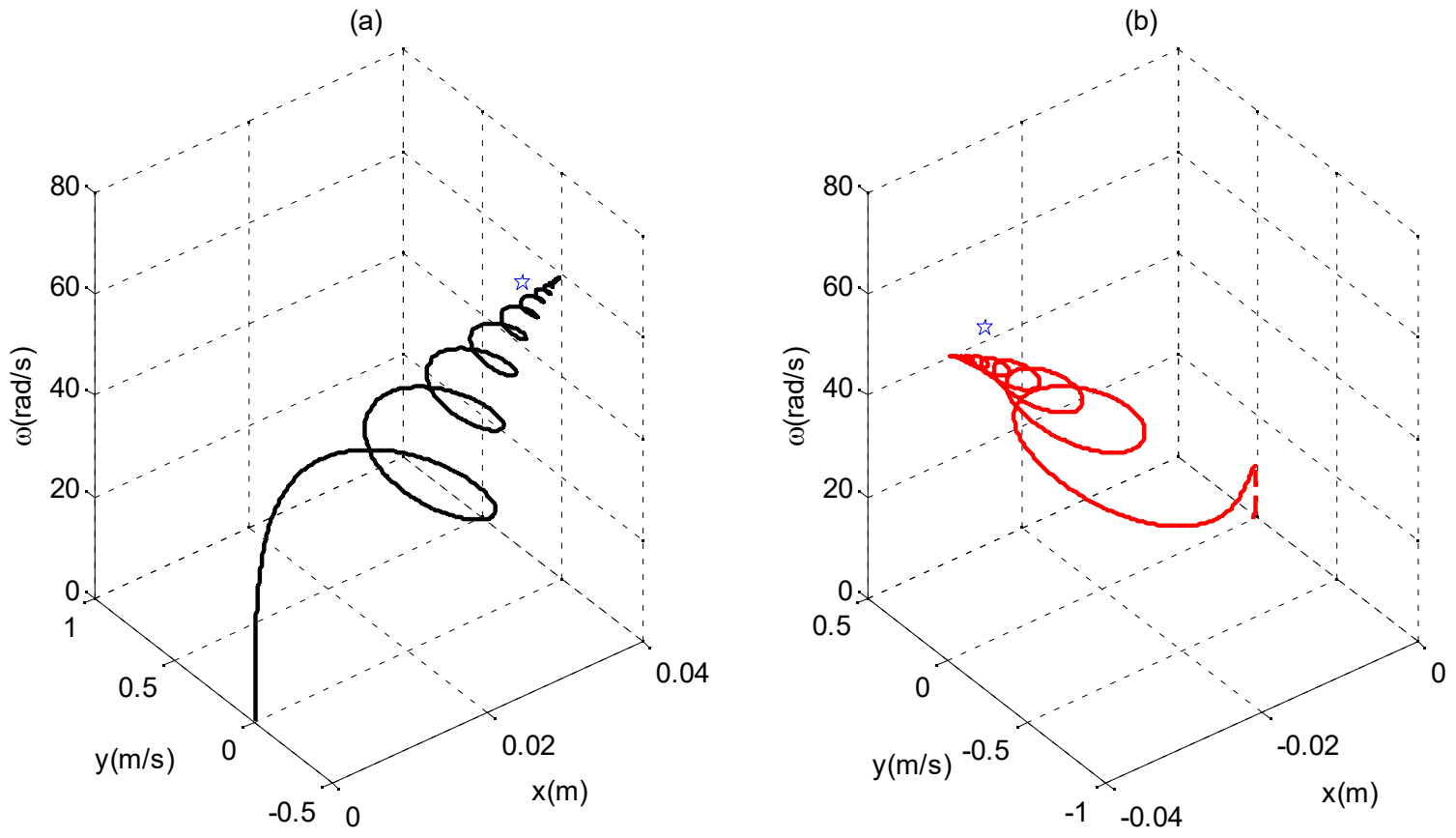

Fig 25.The experimental results from the protype system for $n$=1050r/min. Solid curves and pentagram denote simulated and exprimental results respectivelly. (a)Phasepotrait for inital condition of $E_1$, (b)Phasepotrait for inital condition of $E_3$.

## 6. Conclusions

This paper proposed a novel TCG system and analytically studied the predictive criteria of Melnikov theory for complex dynamical control for both the autonomous and the non-autonomous cases with monostable and bistable states. The new TCG system was designed to show the intrinsic fractal and radical nonlinearities and illustrate spiral chaos. The dynamical response of velocity, restoring force and torque surfaces, equilibrium bifurcation and stability are investigated to are plotted to display the complex dynamical feature of the autonomous system. We show how this archetypal system can transition from monostable ($1<\alpha$) to bistable ($0<\alpha<1$) depending on the geometrical parameter $\alpha$ with the crossing critical value $\alpha$=1, which similarities with the Duffing oscillator. The closed-form three-dimensional Melinkov method, to author knowledge, have firstly made to estimate the homoclinic tangecy of chaotic monition.Finally, we notice that the moddeling method and Melnikov's function discussed here are not limited to the TCG system but can applied also to other system, e.g. energy harvesting, nonlinear circuit, neural network, chemical oscillations, economics. The experimental platform is set up to testified the validity of theoretical result of the proposed TCG model.

**Acknowledgements** This work was supported by the foundation of the National Natural Science Foundation of China (Grant no. 51705241).

**Compliance with ethical standards**

**Funding** National Natural Science Foundation of China (Grant no. 51705241).

**Availability of data and material** These data are collected by the experiment.

**Code availability** The code is written according to the proposed model.

**Conflict of interest** The authors declare that they have noconflict of interest.